\documentclass[useAMS,usenatbib]{mn2e}
\usepackage{graphicx}
\usepackage{amssymb}
\usepackage{lscape}
\usepackage{ulem}
\usepackage{txfonts}

\def\kms{km~s$^{-1}$}

\def\hkpc{$h_{70}^{-1}$ kpc}

\def\aj{AJ}
\def\mnras{MNRAS}
\def\apj{ApJ}
\def\apjl{ApJL}
\def\apjs{ApJS}
\def\aap{A\&A}
\def\msun{M_{\odot}}

\title[Galaxy Pairs in the Sloan Digital Sky Survey III]
{Galaxy Pairs in the Sloan Digital Sky Survey - III: Evidence of Induced Star Formation from Optical Colours.}

\author[Patton et al.] {David R. Patton$^1$, Sara L. Ellison$^2$, 
Luc Simard$^3$,  Alan W. McConnachie$^{3}$, J. Trevor Mendel$^2$\\
$^1$ Department of Physics \& Astronomy, Trent University, 
1600 West Bank Drive, Peterborough, Ontario, K9J 7B8, Canada.\\
$^2$ Department of Physics and Astronomy, University of Victoria, Victoria, British Columbia, V8P 1A1, Canada.\\
$^3$ National Research Council of Canada,
Herzberg Institute of Astrophysics, 5071 West
Saanich Road, Victoria, British Columbia, V9E 2E7, Canada
}

\begin{document}

\date{{\it Accepted for publication in MNRAS on October 27, 2010}.}

\maketitle

\begin{abstract}

We have assembled a large, high quality catalogue of galaxy colours
from the Sloan Digital Sky Survey Data Release 7, 
and have identified 21,347 galaxies in 
pairs spanning a range of projected separations ($r_p < 80$ \hkpc), 
relative velocities ($\Delta v < 10,000$ \kms, which includes projected
pairs that are essential for quality control), and stellar mass ratios 
(from 1:10 to 10:1).  
We find that the red fraction of galaxies in pairs is higher than 
that of a control sample matched in stellar mass and redshift, 
and demonstrate that this difference is likely due to the fact that
galaxy pairs reside in higher density environments than non-paired
galaxies.  We detect clear signs of interaction-induced star formation within 
the blue galaxies in pairs, as evidenced by a higher
fraction of extremely blue galaxies, along with 
blueward offsets between the colours of paired versus control galaxies.  
These signs 
are strongest in close pairs ($r_p < 30$ \hkpc~ and
$\Delta v < 200$ \kms), diminish for more widely separated
pairs ($r_p > 60$ \hkpc~ and $\Delta v < 200$ \kms) 
and disappear for close projected pairs 
($r_p < 30$ \hkpc~ and $\Delta v > 3000$ \kms).
These effects are also stronger in central (fibre) colours than in global 
colours, and are found primarily in low- to medium-density environments.
Conversely, no such trends are seen in red galaxies, apart from
a small reddening at small separations which may result from residual 
errors with photometry in crowded fields. 
When interpreted in conjunction with a simple model of induced starbursts, 
these results are consistent with 
a scenario in which close peri-centre passages trigger induced star 
formation in the centres of galaxies which are sufficiently gas rich, 
after which time the galaxies gradually redden as they separate 
and their starbursts age.

\end{abstract}

\section{Introduction}

Comparisons between galaxy populations throughout 
the redshift range of $0 < z < 1$ indicate that the red sequence has 
roughly doubled in mass during this timeframe, while the 
mass of the blue cloud is unchanged \citep{faber07,martin07,ruhland09}. 
The red sequence gains mass from the quenching of blue galaxies, 
while the corresponding loss in blue cloud mass is balanced by 
the ongoing star formation within blue cloud galaxies.
This evolution is accompanied by an order of magnitude decrease in 
the cosmic star formation rate \citep[e.g.,][]{madau96}, 
a transition from disc-dominated 
to bulge-dominated galaxies \citep{oesch10,lopezsanjuan10}, 
a decrease in the galaxy merger rate \citep[e.g.,][]{lin08}, 
and the hierarchical buildup of massive galaxies.

Galaxy-galaxy interactions and mergers are 
thought to contribute to this evolution by triggering the formation
of new stars, by quenching star formation in gas-rich galaxies, 
and by moving galaxies up the red sequence via dry mergers 
\citep{schiminovich07,dimatteo08,skelton09,wild09}.
Close encounters are also thought
to play a role in producing a wide variety of transient 
astrophysical phenomena, 
such as quasars \citep{hopkins07,treister10,green10}, 
submillimetre galaxies \citep{conselice03,tacconi08}, 
luminous infrared galaxies \citep{wang06,shi09}, 
and ultra-luminous infrared galaxies \citep[ULIRGs; ][]{dasyra08,hou09,chen10}.

\citet{lt78} provided the first clear evidence of enhanced star formation
in interacting galaxies, by comparing the optical colours of morphologically 
peculiar galaxies to normal galaxies.  
In recent years, numerous lines of evidence have confirmed this finding.
The level of enhancement is typically about a factor of two
\citep[e.g.,][]{heiderman09,knapen09,robaina09},
but varies depending on the types of galaxies 
involved (e.g., massive galaxies vs. starforming galaxies), 
how the interactions are identified (e.g., visual classifications 
vs. presence of a close companion), how advanced the interaction/merger is, 
and the method used to measure the star formation 
(e.g., H$\alpha$ equivalent width vs. far infrared emission).
A number of recent studies of close galaxy pairs have demonstrated 
that the enhancement in star formation increases as pair separation 
decreases 
\citep{barton00,lambas03,alonso04,nikolic04,alonso06,geller06,woods06,barton07,
woods07,ellison08,ellison10,woods10}.  
Studying this problem in reverse reveals that galaxies which are undergoing
strong star formation have an increased likelihood of having a close 
companion \citep[e.g.,][]{owers07,li08}.
There is also evidence that recent (rather than ongoing) star formation 
is enhanced in interacting galaxies, from studies of 
E+A galaxies \citep{nolan07,yamauchi08,brown09,pracy09}, 
the absorption-line spectra of early type galaxies \citep{rogers09}, 
and the recent star formation histories of ULIRGs \citep{zaurin10}.

Nevertheless, many questions remain.  Assessment of the 
relative roles of gas-rich versus gas-poor galaxies in 
interactions and mergers has yielded some conflicting results, 
due in part to the methods used to identify these systems and 
detect their star formation.
Interacting galaxies which are identified 
based on morphological signs of interactions are strongly biased towards
systems with large gas fractions, and these tidal disturbances remain 
visible for longer in gas-rich systems \citep{lotz10}.
This issue can be avoided by identifying interacting systems via the presence
of close companions.  However, many such close pair studies employ 
star formation rate (SFR)
indicators which are primarily sensitive to gas-rich galaxies (e.g., those 
which use nebular emission lines); low levels of enhanced star formation in 
gas-poor systems may therefore be overlooked.  
Moreover, these same SFR
indicators are sensitive to relatively 
short-lived {\it ongoing} star formation, 
and therefore may only be able to identify signs of triggered star formation
in systems which are seen very shortly after close passages.
Finally, a perennial problem with studies of interacting galaxies is the
question of ``nature versus nurture''.  That is, if one detects differences
between interacting and non-interacting galaxies, it is difficult to 
distinguish between interaction-induced effects (such as triggered 
star formation) and pre-interaction differences (e.g., if interacting 
and non-interacting galaxies reside in different environments).

One relatively obvious way forward is to analyze the optical colours of 
galaxies in close pairs, and compare them to a fair sample of non-paired
galaxies.  Optical colours can be measured for all types of galaxies, 
and provide a clear method of 
distinguishing between gas rich and gas poor galaxies, due to 
the well-established bi-modality of galaxy colours 
\citep[e.g.,][]{baldry04}.  In addition, induced star formation 
alters the colours of galaxies, and on timescales which are considerably 
longer than those of the starbursts themselves.  
However, while there have been a number
of previous studies of the optical colours of galaxies in interacting/merging 
galaxies, they have yielded some conflicting results.  
The earliest studies of close galaxy pairs generally found that the colours
of galaxies in close pairs are similar to field galaxies 
\citep[see][and references therein]{patton97},
although these studies were limited to small 
samples of galaxy pairs, often without redshifts.
In recent years, large redshift surveys have greatly increased the yield
of close spectroscopic galaxy pairs, allowing differences to emerge.
\citet{depropris05} find that galaxies in 
close pairs are bluer than galaxies in their parent sample.  Other studies 
report an excess of both extremely blue and extremely red galaxies 
in close pairs \citep{alonso06,perez09b} and 
visually identified mergers \citep{darg10}.  
Interpretation of these and other results is complicated by the 
limited size of many close pair samples, deficiencies in the quality 
and size of the control samples, 
uncertainties about the quality of colours measured 
in these crowded systems,  
and the fact that some studies do not directly 
probe colour changes as a function
of pair separation.

Optical colours can also provide 
further insight into the degree to
which triggered star formation is centrally concentrated.  Simulations 
indicate that strong interactions can cause the infall of gas onto the 
central regions of galaxies, triggering star formation 
\citep{mihos94,cox06,dimatteo07}.  
This process may contribute to the 
growth of bulges \citep{barton03,kannappan09,oesch10}, 
even if the interactions do not lead to mergers.
Several lines of observational evidence indicate that 
interaction-induced
star formation tends to be centrally concentrated. 
\citet{bergvall03} and \citet{park09} 
use optical colours to infer that star formation 
is enhanced in the centres of interacting and merging galaxies, 
while \citet{barton00} use both optical colours and H$\alpha$ equivalent 
widths to reach the same conclusion in a sample of close galaxy pairs.
\citet{ellison10} use bulge versus disc colours to 
infer evidence of enhanced star formation in the bulges, 
but not the discs, of galaxies in close pairs.
\citet{rossa07} find that the surface brightness profiles of Toomre-sequence 
galaxies are consistent with the presence of newly formed stars in the 
centres of these merging galaxies, while
\citet{habergham10} find a central excess of core collapse supernovae 
in the cores of disturbed galaxies.
Evidence for the infall of gas onto the centres of galaxies comes from 
an offset of the luminosity-metallicity relation and mass-metallicity  
relation to lower metallicities in close galaxy pairs 
\citep{kewley06,ellison08}, and a higher proportion of strongly disturbed 
systems in lower metallicity galaxies \citep{solalonso10}, 
and is consistent with predictions from the simulations of \citet{rupke10}.
However, there are also clear examples of galaxy-galaxy interactions 
which trigger off-centre star formation \citep{inami10,zhang10} or galaxy wide 
star formation \citep{goto08}.  Furthermore, \citet{knapen09} find no excess of
central star formation in a sample of galaxies with close companions, 
despite the overall SFRs of these galaxies being nearly twice as high as 
galaxies in their control sample. 

We aim to shed new light on these issues by measuring the 
$g-r$ colours of galaxies in Sloan Digital Sky Survey (SDSS) 
close pairs, and comparing them with 
a control sample of non-paired galaxies that are matched 
in both stellar mass and redshift.  In addition, by 
comparing with both wide separation pairs and close projected pairs 
(i.e., interlopers), we wish to tease apart colours differences
which are due to interactions from those which result from environmental
differences or poor photometry.
Finally, by comparing global colours to central (fibre) colours, we 
will investigate the degree to which induced star formation is 
centrally concentrated.
Compared with earlier studies of the colours of galaxies in close pairs, 
our study is unparalleled in terms of the size of the pairs sample, 
the quality of the photometry, the combination of global and central colours, 
the size and robustness of the control sample, the comparison with 
close projected pairs, and the use of colour offsets.

We describe the selection of our pairs and control samples
in Section~\ref{secsample}, along with 
our measurements of global and central colours.  
In Section~\ref{seccolour}, 
we present the overall distributions of global colours in paired galaxies, 
along with their dependence on pair separations and relative velocities.
In Section~\ref{secclass}, we divide our sample into four subsets 
(red sequence, blue cloud, extremely red, and extremely blue), and
explore how the fractions and colours of galaxies in these subsets
depend on projected separation.
We then assess the degree to which these trends are related to central 
(rather than global) colours by analyzing fibre colours
(\S~\ref{secfibre}).  We introduce a new measure called 
colour offset in Section~\ref{secoffset}, and relate our findings to 
predictions from a simple starburst model (Section~\ref{secsb99}). 
We finish with our conclusions in Section~\ref{secconclusions}.
We adopt a concordance cosmology of $\Omega_{\Lambda} = 0.7$, $\Omega_M = 0.3$,
and $H_0 = 70~h_{70}$ \kms Mpc$^{-1}$ throughout the paper.

\section{Sample Selection and Colour Measurements}\label{secsample}

We wish to analyze the optical colours of 
galaxies in close pairs, and to compare them with
a control sample of galaxies which do not have close companions.  
In addition, we will compare close and wide pairs, in order to 
be certain that any effects attributed to ongoing interactions/mergers
decline at wider separations, as would be expected in this scenario.
Finally, we will compare close physical pairs to close 
projected pairs, in order to ensure that our findings are not 
adversely affected by poor photometry due to crowding.
In this section, we describe our initial acquisition of galaxies 
from the SDSS Data Release 7 (hereafter 
DR7) of \citet{dr7}, along with our selection of galaxies in pairs, 
and the creation of an unbiased control sample.

\subsection{A Catalog of SDSS Galaxies}\label{catalog}

Studies of close pairs of galaxies have benefitted greatly from the 
advent of large redshift surveys.  The availability of redshifts for 
both members of a close pair reduces the contamination due to unrelated 
foreground/background companions, and allows one to compare intrinsic 
galaxy properties as a function of projected physical separation and 
relative line-of-sight velocity.  We therefore begin by requiring 
all galaxies in our sample to have secure 
spectroscopic redshifts from the SDSS DR7; 
specifically, we select galaxies from the SpecPhoto table which have 
zConf $>$ 0.7 (i.e., all redshifts are at least 70\% secure).

We further limit our analysis to the SDSS Main Galaxy Sample \citep{strauss02}, 
by requiring extinction-corrected Petrosian apparent magnitudes of 
$m_r$ $\le$ 17.77.  We impose an additional limit of $m_r > 14.5$ in order to
avoid the unreliable deblending of large galaxies, which can  
lead to single galaxies being misclassified as close pairs, triples, etc.
We impose a minimum redshift of 0.01 to ensure that redshifts are primarily 
cosmological, and we impose a maximum redshift of 0.2 
to avoid the regime of high incompleteness and poor spatial resolution.
We also require all objects to be classified 
as galaxies both photometrically (SpecPhoto.Type=3) and spectroscopically 
(SpecPhoto.SpecClass=2).  

Finally, studies of galaxy pairs benefit from knowledge of
the luminosity or mass ratio of every pair. 
We therefore also require every galaxy to be included 
in the MPA-JHU DR7 
stellar mass catalogue\footnote{http://www.mpa-garching.mpg.de/SDSS/}. 
These stellar masses were measured using fits to SDSS $ugriz$
photometry \citep{salim07}, rather than 
using spectral features \citep{kauffmann03}, although in general these
mass estimates agree very well\footnote{See http://www.mpa-garching.mpg.de/SDSS/DR7/mass\_comp.html}.
Together, these criteria yield a catalogue of 615,196 galaxies.  

\subsection{The Pairs Sample}
We identify a sample of galaxy pairs following the general approach of 
\citet{ellison08}.  For each galaxy in the catalogue described above, 
we identify the closest companion satisfying the following criteria: 
\begin{enumerate}
\item projected physical separation of $r_p < 80$ \hkpc
\item line-of-sight rest-frame velocity difference of $\Delta v < 10,000$ \kms
\item stellar mass ratio (companion mass divided by host mass) of 
$0.1 < $ mass ratio $< 10$
\end{enumerate}
If several companions are found for a given host galaxy, the companion with
the smallest $r_p$ is selected.  If no companions are found, the galaxy
is designated as a potential control galaxy (see below).

Finally, following \citet{ellison08}, we randomly remove 67.5\% of galaxies
which are in pairs with angular separations greater than 55$\arcsec$, 
in order to compensate for the fact that pairs with smaller angular separations
would otherwise be underrepresented in our sample.  
This small scale spectroscopic incompleteness results from the well known 
SDSS fibre collision constraint \citep{strauss02}, whereby one cannot 
simultaneously acquire spectra for two galaxies within 55$\arcsec$.  
Thankfully, plenty of these pairs
are nevertheless present in SDSS, due to overlap between adjacent plates 
and the use of two or more plates in some regions.  
\citet{ellison08} use the spectroscopic incompleteness measurements of 
\citet{patton08} to estimate that 67.5\% of pairs at angular separations below
55$\arcsec$ are missed as a result of these fibre collisions.  We therefore 
exclude the same fraction of galaxies in pairs with separations 
$>$ 55~$\arcsec$.

Application of all of these criteria to the catalogue described in
Section~\ref{catalog} yields a sample of
22,777 galaxies with a companion.  Hereafter, we refer to these galaxies as 
``paired galaxies''.  

\subsection{The Control Sample}\label{seccontrol}

In order to ascertain how interactions/mergers affect galaxy properties, 
we wish to compare our sample of paired galaxies with a 
sample of non-interacting galaxies.  However, if one simply 
selects galaxies at random from the remaining (non-paired) galaxies in 
the catalogue, 
the resulting distribution of stellar masses and redshifts 
will be significantly different from the pairs sample 
\citep{ellison08,perez09a}, 
due to the pair selection criteria, SDSS fibre collisions, etc.
Given that many observed properties of galaxies are redshift dependent, 
and that intrinsic galaxy properties are known to correlate 
with stellar mass \citep[e.g.,][]{abbas06},
we therefore create a control sample 
that is matched to the pairs sample in both stellar mass and redshift.
Unlike \citet{ellison10}, we do not attempt to explicitly 
match our control sample to the environment of paired galaxies; however,
we do compare the environments of paired and control galaxies in 
Section~\ref{secrfrac}, and we investigate the dependence of our results
on environment in Section~\ref{secoffset}.

\begin{figure}
\centerline{\rotatebox{0}{\resizebox{8.5cm}{!}
{\includegraphics{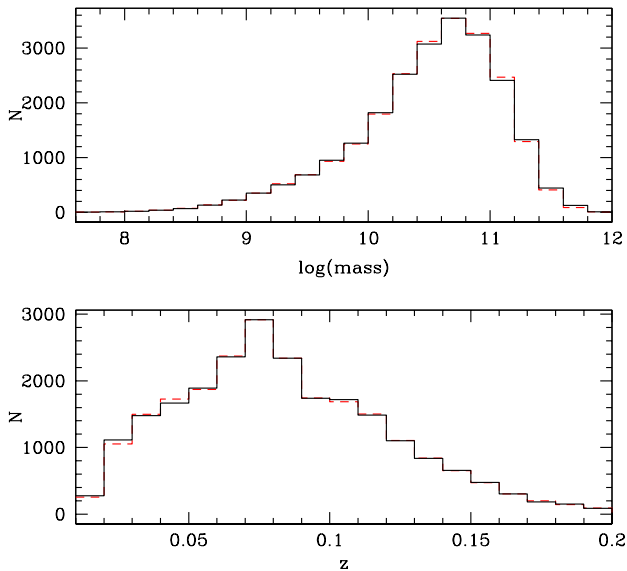}}}}
\caption{Histograms of redshift and stellar mass for galaxies in pairs 
(solid black lines) and in the control sample (dashed red lines) 
are used to assess the success of the control sample matching procedure.  
The control sample histogram has been scaled by a factor of 13, to account 
for the fact that each paired galaxy has 13 associated control galaxies.
K-S test results are 64\% and 68\% for redshift and mass distributions
respectively, indicating that the paired and control galaxies samples are 
consistent with being drawn from the same parent distributions in redshift
and stellar mass.
\label{fighistzm}}
\end{figure}

The control sample is created by first finding the best simultaneous 
match in redshift and stellar mass for each paired galaxy.  This yields
an initial control sample which is equal in size to the paired sample, with 
nearly identical distributions of redshift and stellar mass.  We confirm 
this agreement by carrying out Kolmogorov-Smirnov (K-S) tests on the 
two distributions, in which low significance levels would imply 
significantly different cumulative distribution functions; we instead find 
significance levels of 100.0000\% in each case.  We therefore increase the size 
of our control sample by repeating this exercise until the distributions
in redshift and stellar mass of the control sample start to diverge
significantly from those of the paired sample.  With this approach, 
we are able to create a control sample that is 13 times larger than the
pairs sample.  Figure~\ref{fighistzm} confirms that the resulting redshift and
stellar mass histograms appear to be in 
excellent agreement with one another, and this impression is verified by 
K-S test results of 64\% and 69\% respectively (these values would have
dropped to 11\% and 27\% respectively 
if we had included a fourteenth iteration of the matching procedure).  
The fact that the control 
sample is much larger than the pairs sample means that the control sample will
make a negligible contribution to the statistical errors in our results.

We tag each control galaxy with the properties of the paired galaxy it was 
matched to: specifically, the projected separation $r_p$, the 
rest-frame velocity difference 
$\Delta v$, the stellar mass ratio, and the unique 
identifier (SDSS objID) of the paired galaxy.  This will enable us 
to compare individual paired galaxies with their associated control samples, 
or to compare subsets of the paired sample (selected based on any combination 
of pair properties) with the corresponding control galaxies.  
Both approaches will prove to be very powerful in revealing 
differences in the colours of paired and control galaxies. 

There are several important advantages of this approach.  
The first is that we are able to account
for changes in redshift and/or stellar mass as a function of pair 
separation, $\Delta v$, etc.  
This can be contrasted with other published studies in which 
the properties of control galaxies are averaged over the full sample, 
rather than being computed as a function of pair separation, etc.
Secondly, we can compare any individual paired 
galaxy to its own control sample, thereby providing an additional tool 
for assessing which paired galaxies are most different from their controls.

In the analysis and interpretation that follows later in this paper, 
the reader should 
keep the following two points in mind.  First, any differences between 
galaxies in the paired and control samples cannot be due to 
differences in stellar mass or redshift; therefore, these differences 
may tell us something fundamental about how interacting galaxies 
differ from non-interacting galaxies.
Secondly, if the properties 
of control sample galaxies are found to vary with pair properties 
(e.g., pair separation), this is likely due to 
variation in the stellar mass and/or redshift distribution of 
galaxies in the pairs sample.  While both types of information are 
in principle useful, the latter must be treated with caution, 
since changes in the stellar mass and/or 
redshift distribution may be the result of 
selection effects in the pairs sample (e.g., redshift-dependent 
selection effects, incompleteness in the stellar mass catalog, 
spectroscopic incompleteness, etc.).  As a result, we will focus 
primarily on differences between the pairs and control samples.

\subsection{{\sc GIM2D} Fits}\label{secgim2d}

In addition to satisfying the basic requirements of our pair and 
control sample algorithms, we now further require that all galaxies 
in our sample have high quality global (integrated) rest-frame 
$g-r$ colours measured by \citet{simard10}.
These colours were computed using the Galaxy Image 
2D ({\sc GIM2D}) software of \citet{simard02}, 
using simultaneous $g$ and $r$ band fits to the SDSS images.
All colours are corrected for Galactic extinction, and converted 
to rest-frame quantities using the $k$-correction 
software of \citet{blanton07}.
\citet{simard10} demonstrate that these fits, 
which were carried out using improved  
background subtractions and segmentation maps, provide robust  
colour measurements for galaxies which have close companions.  
This allows us to avoid known challenges with the photometry of 
crowded systems \citep[e.g.,][]{patton05,masjedi06,depropris07}\footnote{We 
investigate the importance of careful photometry in crowded fields 
in Section~\ref{seceberphot}.}.

In particular, we require each galaxy in the pairs
and control sample to satisfy the following criteria:
\begin{enumerate}
\item successful {\sc GIM2D} simultaneous $g$- and $r$-band fit 
from \citet{simard10}
\item rest-frame $g-r$ colour error less than 0.1 mag
\item the fibre colour predicted by the {\sc GIM2D} model fit must be
within 0.1 mag of the observed SDSS fibre 
colour\footnote{More precisely, 
the ``$\Delta$ (fibre colour)'' parameter defined and reported in 
\citet{simard10} must be within 0.1 mag of 
the control sample mean.}
\item visual inspection confirms that the object is a distinct galaxy
(inspection is complete only for pairs with $r_p < 10$ \hkpc) 
\end{enumerate}

Overall, 94\% of galaxies in our preliminary paired and control galaxy 
samples satisfy all of these criteria.  This yields final samples of 
21,347 paired galaxies and 261,023 control 
galaxies\footnote{In cases where a paired galaxy has been removed from the 
sample due to application of the above criteria, its associated control 
galaxies are also removed.},
with an average of 
12.23 control galaxies for each paired galaxy.  We note that these samples
are substantially larger than those of \citet{ellison10}, since their pairs
sample is restricted to DR4 galaxies with $z < 0.1$, and their control sample 
is only four times larger than their pairs sample (due to their additional 
requirement for a match on local density).

\subsection{SDSS Fibre Colours}\label{secfibredata}

The \citet{simard10} catalogue 
contains global colours for each galaxy in
our paired and control galaxy samples, 
measured using bulge+disc decomposition.  These colours are representative 
of the galaxies as a whole, and should be sensitive to induced star formation 
that is either global or centrally concentrated.  However, the SDSS database
also provides measurements of fibre colours, which are measured 
within the central 3 arcseconds of each galaxy.  We have converted these
to rest-frame fibre colours using the extinction corrections from 
the SDSS database and the $k-$corrections described in the preceding section.
These fibre 
colours provide a probe of the central star formation in each galaxy, 
although the degree to which these
colours are non-global depends on the covering fraction (CF) of the fibres.
Moreover, if we wish to compare the fibre colours of paired and control 
galaxies, we must ensure that they have similar covering fraction 
distributions.

\begin{figure}
\rotatebox{0}{\resizebox{8.5cm}{!}
{\includegraphics{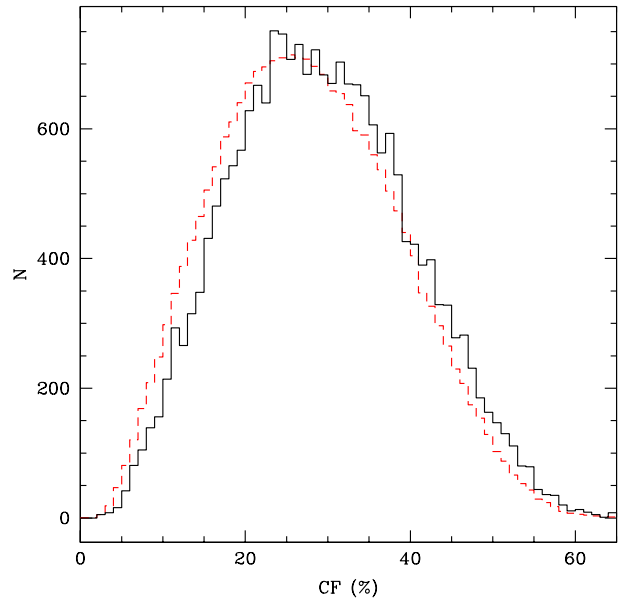}}}
\caption{Histograms of $r-$band 
covering fraction (CF) are shown for paired galaxies
(solid black lines) and the control sample (dashed red lines).
These distributions are seen to be broadly similar, despite the fact that
paired and control galaxy were not matched using covering fractions.  
The offset towards larger CF's for paired galaxies, 
which is on the order of 1.5\%, is sufficiently small that 
it should have a negligible effect on the resulting fibre colours.
\label{fighistcf}}
\end{figure}

In Figure~\ref{fighistcf}, we present $r-$band 
CF histograms of paired and control 
galaxies.  We find broadly similar distributions, with a small offset 
($\sim 1.5\%$) towards larger CF's for paired galaxies.  
While these samples were not 
matched on CF, reasonable agreement is to be expected, given that each 
paired galaxy is matched in both redshift and stellar mass to its 
control galaxies\footnote{That is, for a given galaxy, the CF depends on 
its distance and its size; redshift and stellar mass provide a proxy for 
these quantities.}.  
The offset is in the direction 
expected, given that \citet{ellison10} report that galaxies in a subset of
this pairs sample have somewhat higher bulge fractions than control galaxies.
Most importantly, the offset in CF 
between paired and control galaxies 
is sufficiently small that it should have a negligible influence on 
the resulting colours, allowing us to 
make a fair comparison of fibre colours between 
paired and control sample galaxies.  

We also note that the median CF of paired galaxies 
is 29\%, and 96\% of paired galaxies have covering fractions of less than 
50\%.  Therefore, these fibre colours do in fact provide a good probe
of central colours.
Moreover, unlike the global colours of \citet{simard10}, 
the fibre colours are model independent, and
they are not affected by deblending problems due to 
close neighbouring galaxies or stars.  
Therefore, fibre colours also provide an independent 
check on the trends reported in this study.

\section{Analysis of Global Colours}\label{seccolour}

Armed with secure measurements of global and fibre colours,  
we now set out to find and understand any differences 
between the colours of paired and control galaxies.

\subsection{The Dependence of Mean Colours on $\Delta v$}\label{secgrdv}

Our paired galaxy sample spans a line of sight rest-frame velocity difference
of $0 < \Delta v < 10,000$ \kms.  Pairs with $\Delta v > 3000$ \kms~ 
cannot be physically close together, as the required peculiar velocities 
for this scenario are unphysical.  We therefore will refer to these systems
as {\it projected pairs} (interlopers), 
and will use them as a sanity check; specifically, 
we would expect any real differences between paired and control galaxies
to disappear in the projected pair sample.

\begin{figure}
\rotatebox{0}{\resizebox{8.5cm}{!}
{\includegraphics{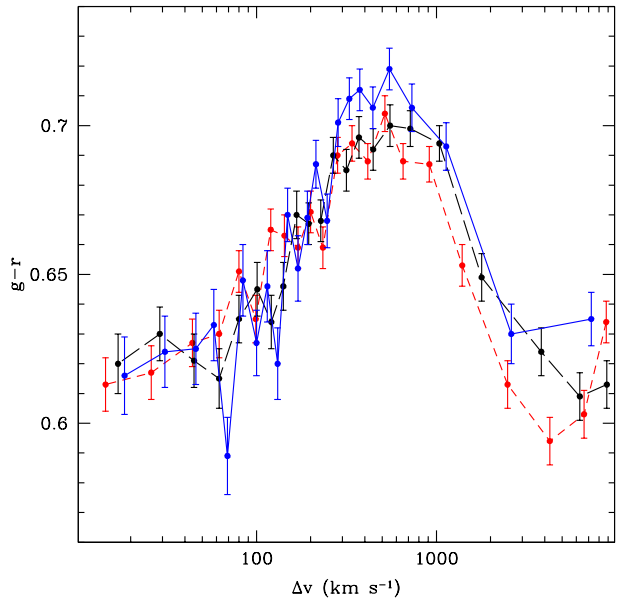}}}
\caption{Mean {\sc GIM2D} global colours 
are plotted versus rest-frame velocity difference ($\Delta v$) for three 
subsets of the pair sample: $r_p < 30$ \hkpc~ (blue symbols; solid lines), 
$30 < r_p < 55$ \hkpc~ (black symbols; long dashed lines), 
and $55 < r_p < 80$ \hkpc~ (red symbols; short dashed lines).  
Error bars in this figure and elsewhere in the paper refer to the 
standard error in the mean, unless otherwise specified.
Mean colours are reddest at $300 \lesssim \Delta v \lesssim 1200$ \kms, 
presumably due to the fact that these pairs lie in the 
highest density environments.  
\label{figgrdv}}
\end{figure}

In order to focus on the effects of galaxy interactions, we wish to impose 
a maximum $\Delta v$ on our non-projected pairs sample.  The primary motivation 
is to minimize chance superposition of non-interacting galaxies within 
relatively dense environments such as groups or clusters.  
\citet{ellison10} show that there is a correlation between 
$\Delta v$ and projected galaxy density, in that pairs with higher $\Delta v$
lie in regions of higher density 
(within their sample of pairs with $\Delta v <$ 500 \kms).  
To guide our choice of a maximum $\Delta v$, we plot in Figure~\ref{figgrdv}
the mean global 
colours of galaxies in pairs as a function of $\Delta v$, considering
three subsets of the pairs sample: close pairs ($r_p < 30$ \hkpc), 
intermediate separation pairs ($30 < r_p < 55$ \hkpc) and wide pairs
($55 < r_p < 80$ \hkpc).  

We find that pairs with $300 \lesssim \Delta v \lesssim 1200$ \kms~ 
have mean colours which are quite red with respect to both low and 
high velocity pairs.  
This is true for close, intermediate, and wide 
separation pairs, indicating that this trend is unlikely to be 
associated with galaxy interactions/mergers.  
Instead, we interpret these redder colours as being due to the 
higher densities probed by these relative velocities \citep{ellison10}, 
and the fact that 
galaxy colour and local density are correlated.  This is consistent with 
the relatively high proportion of late type galaxies in low velocity pairs 
($\Delta v < 200$ \kms) reported by \cite{park09}.
While these higher velocity pairs are certain to include some ongoing 
interactions and eventual mergers (and in fact clear signs of interactions 
are seen in the images of some systems), 
we elect to avoid the expected high superposition rate in this regime 
by imposing a maximum $\Delta v$ of 200 \kms, which 
comfortably avoids these higher velocity environments 
while allowing us to retain 
a sizeable sample of pairs.  For comparison, other close pair studies have 
imposed less restrictive $\Delta v$ limits ranging from 
$\Delta v = 350$ \kms~ \citep{lambas03,alonso04,alonso06,perez09b}, 
to 500 \kms~ 
\citep{patton00,patton02,depropris05,woods07,ellison08,ellison10,woods10} 
and up to 1000 \kms~ \citep{barton00,geller06,barton07}.  

\subsection{The Dependence of Mean Colours on $r_p$}\label{secgrrp}

\begin{figure}
\rotatebox{0}{\resizebox{8.5cm}{!}
{\includegraphics{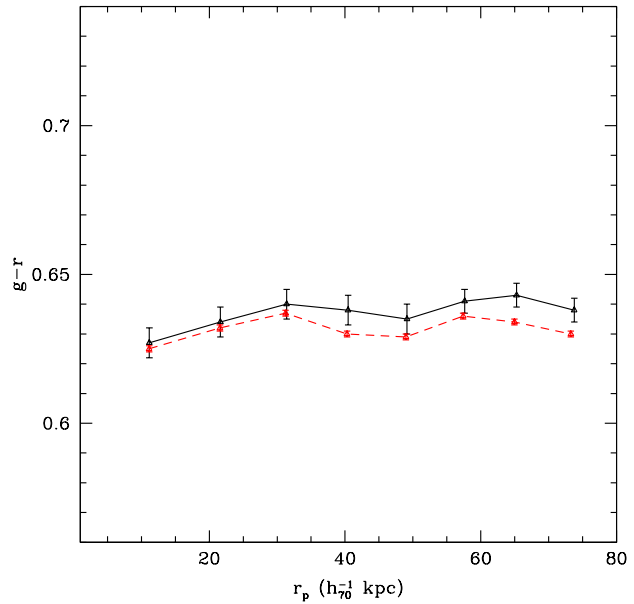}}}
\caption{Mean {\sc GIM2D} global colours are plotted versus projected 
separation ($r_p$) for paired galaxies (black symbols and solid line) and 
control galaxies (red symbols and dashed line).  
The sample is restricted to $\Delta v <$ 200 
\kms.  The vertical scale is the same as in Figure~\ref{figgrdv}, thereby 
emphasizing that mean colours have a much stronger dependence on $\Delta v$ 
than on $r_p$.  The mean colours of paired galaxies  
decrease slightly towards small pair separations (by $\sim$ 0.01 mag), 
whereas their associated control samples have mean colours that are 
relatively constant with respect to $r_p$.
\label{figgrmeannew}}
\end{figure}

We now proceed to compare the colours of galaxies in 
low velocity pairs ($\Delta v < 200$ \kms) with their control galaxies. 
In Figure~\ref{figgrmeannew}, we plot 
mean global colours as a function of $r_p$, for both 
paired and control galaxies.  Compared with the strong dependence of 
mean colour on $\Delta v$ that was seen in Figure~\ref{figgrdv}, 
we find that the global colours of low velocity pairs 
vary much less with $r_p$.
The mean colour of galaxies in pairs decreases smoothly as pair separation 
decreases, with galaxies in the closest pairs being on average about 0.01 mag 
bluer than galaxies in the widest pairs.  
No significant change in the mean colours of the associated control galaxies
is seen over this range in $r_p$.  The closest pairs have global colours
which are equivalent to their controls, whereas the widest pairs are 
$\sim 0.01$ mag redder on average.  

\subsection{The Distribution of $g-r$ Colours}\label{secgrhist}

Based on the lack of a dependence of mean colour on $r_p$, 
one might be tempted to conclude that (a) the colours of 
galaxies are generally unaffected by galaxy-galaxy interactions or (b) that 
few of the galaxies in close pairs are actually undergoing interactions.
However, as has been shown elsewhere \citep[e.g.,][]{balogh04}, 
{\it mean} colours alone 
are not very sensitive to changes within galaxy populations.  For example, 
if there are excesses of both extremely red and extremely blue 
galaxies within samples of interacting/merging galaxies, 
as has been reported elsewhere 
\citep[e.g.,][]{alonso06,darg10}, 
these effects might cancel out, at least in part, when computing mean colours.  

\begin{figure}
\rotatebox{0}{\resizebox{8.5cm}{!}
{\includegraphics{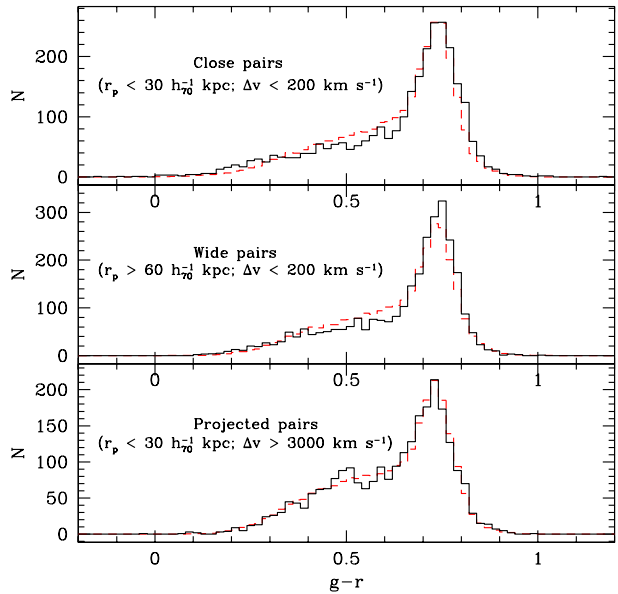}}}
\caption{Histograms of global $g-r$ are shown for pairs (solid black lines) 
and their associated control galaxies (dashed red lines). 
The lower plot is for projected pairs  
($r_p < 30$ \hkpc~ and $3000 < \Delta v < 10,000$ \kms), the middle plot 
is for wide pairs ($r_p > 60$ \hkpc~ and $\Delta v < 200$ \kms), and 
the upper plot is for close pairs ($r_p < 30$ \hkpc~ and $\Delta v < 200$ \kms).
\label{figgrhist}}
\end{figure}

With this in mind, we now compare the colour distributions of paired 
and control galaxies.  We begin by plotting 
histograms of global colours in 
Figure~\ref{figgrhist}, for 
close pairs, wide pairs, and projected pairs.  
Overall, the distributions of paired galaxy colours are broadly 
similar to those of the associated controls, with a distinct 
red sequence and a more extended distribution of blue galaxies 
(the ``blue cloud'').  This is the aforementioned colour bimodality, 
within which the relative proportion of red versus blue galaxies 
has been found to depend on environment and luminosity \citep{balogh04}.
However, we find a small but significant deficit in 
galaxies with intermediate global colours ($0.4 < g-r < 0.65$) for
galaxies in close and wide pairs with 
$\Delta v <$ 200 \kms~ (relative to their control samples), and an
excess of galaxies which are relatively red (on the redward half of the 
red sequence peak)\footnote{We will revisit this excess of red galaxies in 
Section~\ref{secrfrac}}.  
These differences are not seen in the projected pairs sample.  
In addition, a small but significant population of extremely blue galaxies 
($g-r \lesssim 0.3$) is seen in the close pairs sample, but is 
greatly diminished in the wide pairs sample, 
and non-existent in the projected pairs sample.  This lends support 
to the notion that these extremely blue galaxies may be directly associated 
with galaxy-galaxy interactions.  We will return to this intriguing 
sub-population in Section~\ref{seceber}.

\section{Colour Classifications}\label{secclass}

\begin{figure}
\rotatebox{0}{\resizebox{8.5cm}{!}
{\includegraphics{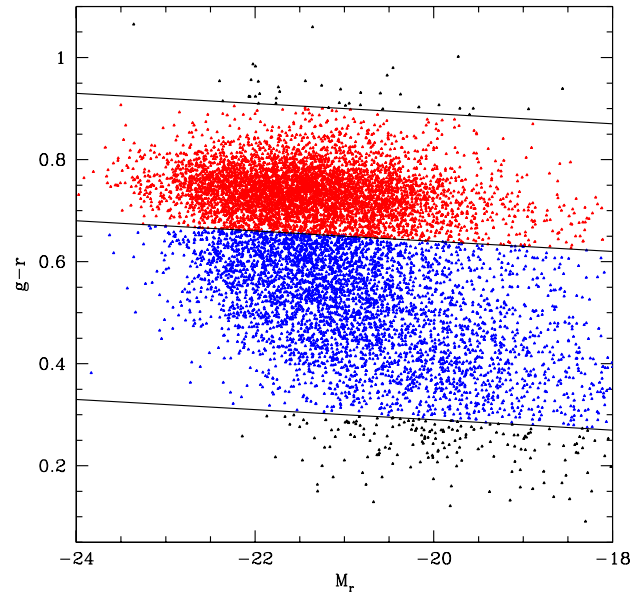}}}
\caption{
The colour magnitude diagram of 10000 galaxies randomly selected 
from the control sample is used 
to illustrate the division our sample into four different colour categories.
The categories are: extremely red (black symbols above upper line), 
red sequence (red symbols), 
blue cloud (blue symbols), 
and extremely blue (black symbols below lower line).
The three solid lines separate these subsets, and 
intersect $M_r = -21$ at $g-r =$ 0.9, 0.65, and 0.3 (top to bottom).
Each line has a slope of $-0.01$, which provides a good fit to the observed
slope of the red sequence.  
\label{figcmdcat}}
\end{figure}

In order to further examine these differences between paired and control
sample galaxies, we now 
divide the sample into four subsets based on colour and 
absolute magnitude.  Figure~\ref{figcmdcat} illustrates this division of the 
sample into extremely red galaxies, red sequence galaxies, 
blue cloud galaxies, and extremely blue galaxies.  
The division between red sequence and blue cloud galaxies corresponds 
to a line with slope $-0.01$ which passes through 
$g-r$ = 0.65 at $M_r = -21$.  The slope provides a good fit to the 
colour magnitude relation seen on Figure~\ref{figcmdcat}, and the 
intercept was selected by examining the 
colour histograms of Figure~\ref{figgrhist}.  Throughout the remainder 
of this paper, this division is used to distinguish between red and 
blue galaxies.

We also identify subsets of extremely red and extremely blue galaxies.
The criterion of $g-r >$ 0.9 at $M_r = -21$ for extremely red galaxies 
applies to the reddest 1\% of galaxies in projected pairs,
and is sufficiently red that galaxies
are unlikely to have been scattered from the red sequence 
(recall from \S~\ref{secgim2d} that all galaxies are 
required to have $g-r$ colour errors of $<$ 0.1 mag). 
This threshold is slightly less strict than 
the $g-r$ = 0.95 cut used by \citet{alonso06}. 
Our threshold for extremely blue galaxies corresponds to $g-r$ = 0.3 at 
$M_r = -21$, and applies to the bluest $\sim$1\% of galaxies in the 
projected pairs sample. 
This threshold is notably stricter 
than the extremely blue cut of $g-r$ = 0.4 employed by \citet{alonso06}.
\citet{west09} find that rising SFRs are needed to produce colours
bluer than $g-r$ = 0.3.

\subsection{The Red Fraction}\label{secrfrac}

\begin{figure}
\rotatebox{0}{\resizebox{8.5cm}{!}
{\includegraphics{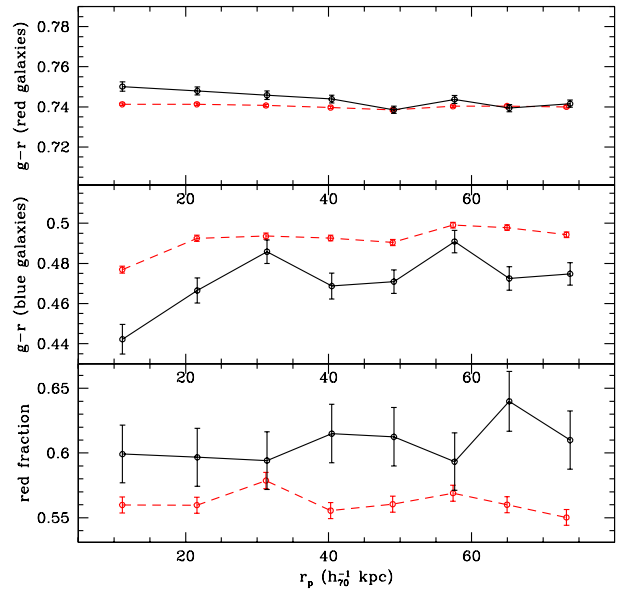}}}
\caption{Trends in the global colours of blue and red galaxies with 
projected separation $r_p$ are investigated.  The lower plot gives the 
red fraction (the fraction of galaxies which are classified (see 
Figure~\ref{figcmdcat}) as red sequence
or extremely red) for paired galaxies (black symbols; solid lines) 
and their associated 
control galaxies (red symbols; dashed lines).
The middle plot gives the mean colour of blue galaxies (those classified as
blue cloud or extremely blue), and the upper plot gives the mean colour of 
red galaxies (those classified as red sequence or extremely red).
\label{figbluered}}
\end{figure}

The fraction of galaxies which are classified as red (either red sequence or 
extremely red), hereafter called the red fraction, is plotted versus 
projected separation in the lower panel of Figure~\ref{figbluered}.
The red fraction of paired galaxies is consistently larger than the
associated control sample at all separations, although this difference
may decline at smaller separations.  These findings are consistent with the 
dependence of mean colours on separation reported in Section~\ref{secgrrp}, 
and with the excess of red sequence galaxies (and corresponding deficit of 
blue cloud galaxies) described in Section~\ref{secgrhist}.  

We note that other studies 
have also reported that galaxies in pairs are significantly redder than 
galaxies without nearby companions \citep[e.g.,][]{perez09b}, 
with correspondingly higher bulge fractions than their isolated 
counterparts \citep{deng08,ellison10}.
The most obvious cause of this difference would be
if pairs reside in higher density environments, since the red fraction 
is known to increase with density \citep{balogh04,baldry06,cooper06}. 
\citet{lin10} show that gas-poor pairs reside preferentially in higher 
density environments, and that this is due primarily to the colour-density
relation.
\citet{barton07} find that paired galaxies in simulations 
occupy higher-mass haloes than isolated galaxies, and predict that
this should lead to mean $g-r$ colours which are about 0.05 mag redder 
than field galaxies.  The fact that we find a smaller difference than this 
($\lesssim 0.01$ mag in the mean; \S~\ref{secgrrp})
is likely due to the fact that our control sample is matched to the 
pairs in both stellar mass and redshift, thereby providing a fairer comparison
than random field galaxies.  

Nevertheless, we can test this hypothesis directly by 
using the \citet{baldry06} measurements of projected local density ($\Sigma$)
as a probe of environment.  $\Sigma$ is computed using the distances to 
the fourth and fifth nearest neighbours within 1000 \kms.
These measurements have recently been updated 
to include DR7 galaxies.  The suitability of these measurements for close pair 
studies is addressed by \citet{ellison09,ellison10}.  
The \citet{baldry06} requirement for redshifts to lie 
within the range 0.010-0.085 means that these measurements are available 
for only 57\% of the galaxies in our full pairs sample.  However, as 
our control sample is matched in redshift to the pair sample, we are 
still able to make a fair comparison between the local densities of paired
and control galaxies.

\begin{figure}
\rotatebox{0}{\resizebox{8.5cm}{!}
{\includegraphics{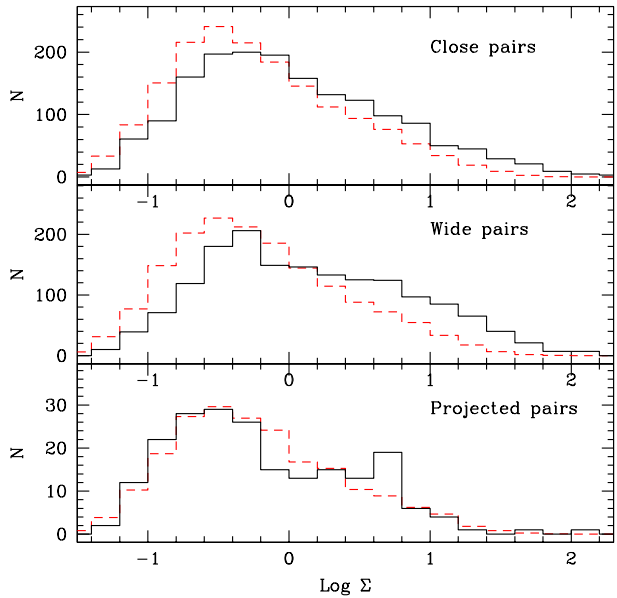}}}
\caption{Histograms of projected local density $\Sigma$ 
are shown for pairs (solid black lines) 
and their associated control galaxies (dashed red lines). 
The lower plot is for projected pairs  
($r_p < 30$ \hkpc~ and $3000 < \Delta v < 10,000$ \kms), the middle plot 
is for wide pairs ($r_p > 60$ \hkpc~ and $\Delta v < 200$ \kms), and 
the upper plot is for close pairs ($r_p < 30$ \hkpc~ and $\Delta v < 200$ \kms).
\label{figdensity}}
\end{figure}

In Figure~\ref{figdensity}, we provide histograms of 
$\Sigma$ for paired and control galaxies.  As with Figure~\ref{figgrhist}, 
we separate our pairs sample into close pairs, wide pairs, and projected pairs.
We find that galaxies in close and wide pairs are skewed to higher 
local densities than their associated control galaxies, with this difference 
being nearly two times larger in wide pairs than in close pairs.
No significant difference is seen in the $\Sigma$ distributions 
of projected pairs
and their control galaxies.  These findings are consistent with the 
hypothesis that the higher fraction of red galaxies in close and wide 
pairs is due to these systems residing in higher density environments 
than their control galaxies, and the inference that more of these 
pairs result from chance superpositions in overdense regions 
\citep{alonso04,perez06}.  Moreover, the fact that this excess is 
more pronounced for wide pairs than close pairs is consistent with 
the expectation that chance superpositions within groups and clusters 
should be more common in wide pairs than in close pairs \citep{alonso04,lin10}.

\subsection{The Colours of Red and Blue Galaxies}

We now proceed to assess the colours of red and blue galaxies separately.
The middle and upper panels of 
Figure~\ref{figbluered} show the mean colours of galaxies classified
as blue (blue cloud or extremely blue) and red (red sequence or extremely
red) respectively.  
In both cases we see clear differences between paired and control galaxies.
Blue galaxies in pairs are bluer than blue galaxies in the 
control sample at all separations, 
with a typical offset decreasing from $\sim$ 0.03 mag for close pairs 
($r_p < 30$ \hkpc) to 0.02 mag for wide pairs ($r_p > 60$ \hkpc).
In contrast, red galaxies in pairs are {\it redder} than red control galaxies 
by $\sim$ 0.01 mag at the closest separations, with this 
relatively small offset becoming negligible beyond $\sim$ 25 \hkpc.
The fact that these trends go in opposite directions (blueward and 
redward) means that they will reduce any associated trends 
in mean colour or red fraction, which is consistent with our findings
in Sections~\ref{secgrrp} and~\ref{secrfrac}.  

\subsection{Extremely Blue and Extremely Red Galaxies}\label{seceber}
In Section~\ref{secgrhist} and Figure~\ref{figgrhist}, we noted a population
of extremely blue galaxies which are present in the close pair sample but
nearly absent in the wide pair sample (and 
non-existent in the projected pair sample).
This is reminiscent of the excess populations of extremely blue galaxies
reported in several close pair studies \citep{alonso06,perez09b,darg10}, 
although we do not detect an obvious population of
of extremely red galaxies as found in these same studies.

\begin{figure}
\rotatebox{0}{\resizebox{8.5cm}{!}
{\includegraphics{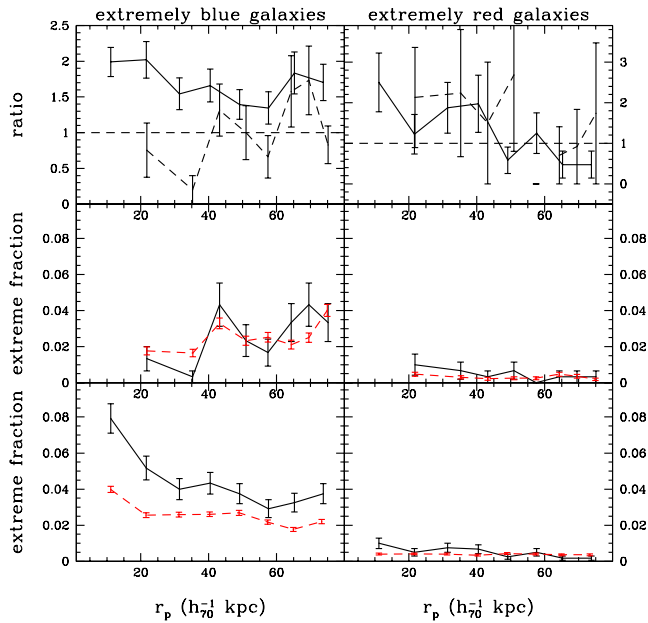}}}
\caption{The prevalence of extremely blue (red) galaxies is
explored using three plots in the lefthand (righthand) column.
The lower panels report the fraction of galaxies which are 
extremely blue or red for low velocity ($\Delta v < 200$ \kms) pairs, 
whereas the middle panels refer to projected pairs 
($3000 < \Delta v < 10,000$ \kms).  
In the lower and middle plots, black symbols and solid lines refer to 
paired galaxies, while red symbols and dashed lines refer to control galaxies.
The upper panels report the ratio of extremely blue/red fractions for
pairs vs. control for low velocity pairs (solid black lines) and
projected pairs (dashed black lines).  The black horizontal dashed line in 
the upper panel corresponds to a ratio of one (i.e., no difference
between pairs and control).  
All error bars in this plot refer to Poisson errors.
\label{figeber}}
\end{figure}

To explore this further, we first 
plot the fraction of galaxies in our $\Delta v < 200$ \kms~ pairs 
which are extremely blue in the lower left panel of Figure~\ref{figeber}.  
We find a clear excess of extremely blue galaxies, rising from 
$\sim$ 3\% at wide separations to $\sim$ 7\% at small separations.  
A similar but less pronounced trend is seen in the control sample, 
implying that part of the trend in paired galaxies is due to a 
change in the mix of stellar mass and/or redshift with separation.  
We repeat these measurements for our sample of projected pairs 
($3000 < \Delta v < 10,000$ \kms) in the middle left panel of 
Figure~\ref{figeber}, finding no rise as $r_p$ decreases\footnote{We note that
there are far fewer projected pairs ($3,000 < \Delta v < 10,000$ \kms) than 
low velocity pairs ($\Delta v <$ 200 \kms),
since correlated pairs are much more common than chance superpositions.
This explains why the error bars on the projected pairs sample are larger, and 
why these data do not extend as far inwards in $r_p$.}.
This essentially rules out the 
possibility that the rise towards small $r_p$ seen for true close pairs  
could be due to poor photometry in crowded pairs.
Finally, we also compute the ratio of the extremely blue fractions in pairs 
versus control (upper left hand panel of Figure~\ref{figeber}), 
in order to see how the rise in the extremely blue fraction 
of pairs compares with the rise for the control sample.  
Contrary to what is 
observed for projected pairs (dashed line), the ratio for low velocity pairs 
(solid line) is significantly higher than unity throughout the full range of 
$r_p$, rising from about 1.5x for wide pairs to about 2x for the 
closest pairs.

In the right hand panels of Figure~\ref{figeber}, we apply the same approach 
to the extremely red galaxies.  We find a much lower fraction of galaxies
in this category, rising from $\sim 0.3\%$ for wide pairs to $\sim 1\%$ 
for the closest pairs.  
The control sample remains constant at about $0.4\%$ for all separations.
The ratio of pairs to control exhibits a gradual rise in 
extremely red galaxies towards the smallest pair separations, reaching 
a factor of 2.5x at $r_p \sim 10$ \hkpc.  However, a comparable rise is 
also seen in the projected pairs sample, leading us to believe that this 
rise may be due to residual problems with the photometry of crowded systems.

Together, our results are qualitatively consistent with the 
high fractions of extremely blue galaxies in pairs reported by 
\citet{alonso06}, \citet{perez09b}, and \citet{darg10}, 
but appear to be at odds with the substantial fractions
of extremely red galaxies described in these same studies.  
For example, \citet{alonso06} find extremely 
red fractions ranging from $\sim 7$ to $16\%$ for three different 
environmental classes, using a {\it stricter} criterion ($g-r > 0.95$) 
than we do ($g-r > 0.9$ at $M_r = -21$).  

\subsection{The Impact of Photometric Quality on Extremely Red/Blue 
Galaxies}\label{seceberphot}

We have found that a substantial number of galaxies in pairs are 
classified as extremely blue (up to $\sim 7\%$ in the closest pairs), 
whereas very few are classified as extremely red ($\lesssim 1\%$ 
in the closest pairs).  The widest pairs also contain roughly 50\% 
more extremely blue galaxies than the control sample, but no 
excess of extremely red galaxies.  Moreover, the trends seen in the 
extremely blue (red) fraction are absent (present) in the projected pairs 
sample.  This implies that poor photometry cannot explain the extremely 
blue population, but may explain the extremely red population.  

\begin{figure}
\rotatebox{0}{\resizebox{8.5cm}{!}
{\includegraphics{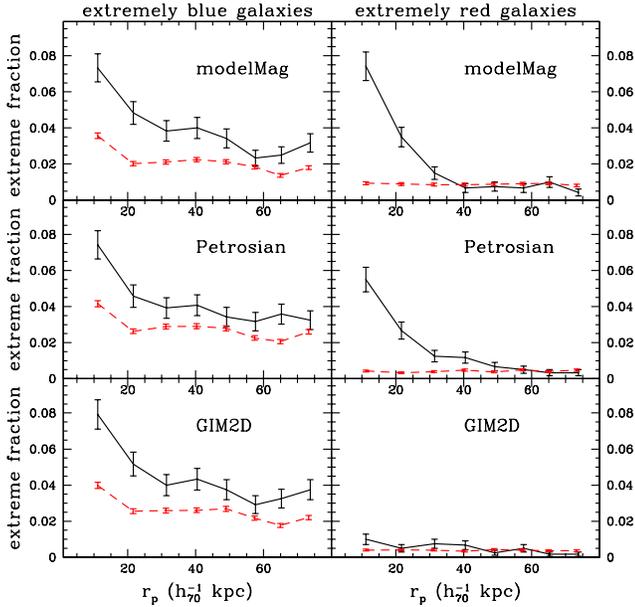}}}
\caption{The fraction of extremely blue (red) galaxies 
is plotted  in the lefthand (righthand) column for three different indices
of $g-r$ colours.  
The lower plots reports the global {\sc GIM2D} colours used in our analysis.
The middle and top panels instead employ SDSS-measured 
Petrosian and modelMag colours respectively, for the same galaxies.
Pairs are displayed using black symbols and solid lines, 
while the associated control galaxies
are displayed using red symbols and dashed lines.
All error bars in this plot refer to Poisson errors.
\label{figebercomp}}
\end{figure}

\citet{simard10} provide a clear demonstration that the standard 
SDSS pipeline does a poor job of galaxy photometry for closely 
separated pairs, whereas their recomputed {\sc GIM2D} global colours are 
much more secure.  To directly assess 
the effects of using these different measurements of $g-r$ colours, 
we compute the extremely blue and extremely red fractions 
in Figure~\ref{figebercomp}, 
using our {\sc GIM2D} global colours (bottom row), 
SDSS Petrosian colours (middle row), and SDSS modelMag colours (top row).  
Exactly the same set of paired and control galaxies is used in each case.
The rise in the extremely blue fraction of close pairs towards small 
separations is seen with all three colour indices.  Conversely, 
a large increase in the extremely red fraction of close pairs towards
small separations is seen only with SDSS Petrosian and modelMag colours, 
reaching 6\% and 8\% respectively.  

This provides compelling evidence 
that poor photometry is in fact largely responsible for the large extremely red 
fractions seen in close pairs when using photometry directly from SDSS, 
and that the re-computed colours used in our analysis are effective 
in removing nearly all of these anomalously red systems.  Related factors 
which may contribute to the lower extremely red fractions found in our 
study include our imposition of a bright apparent magnitude limit of
$m_r > 14.5$ (which allows us to avoid the brightest galaxies where 
deblending is particularly problematic) and our use of a lower relative velocity
threshold than most other studies (which preferentially avoids pairs 
in more crowded regions).

We caution 
that the small rise in our {\sc GIM2D} extremely red fraction at
small $r_p$ 
could be due to residual effects from poor photometry in crowded systems.
More importantly, it seems likely that published reports of large fractions of
extremely red galaxies in close pairs or merging galaxies
\citep[e.g.,][]{alonso06,perez09b,darg10} are the result of poor SDSS 
photometry, rather than dust obscuration or other physical effects 
associated with induced star formation.

\begin{figure*}
\rotatebox{0}{\resizebox{15cm}{!}
{\includegraphics{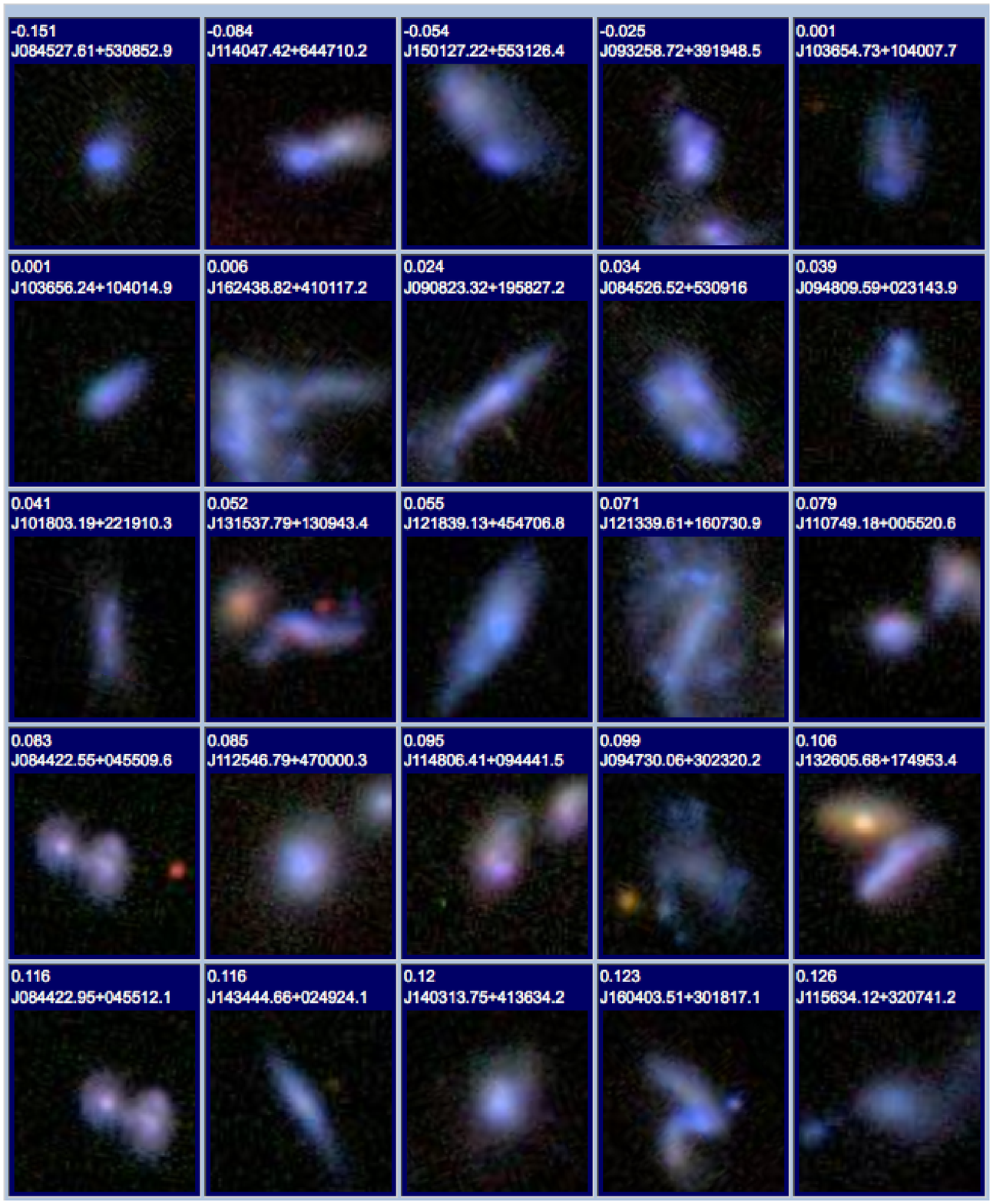}~~~~~~~~~~~~\includegraphics{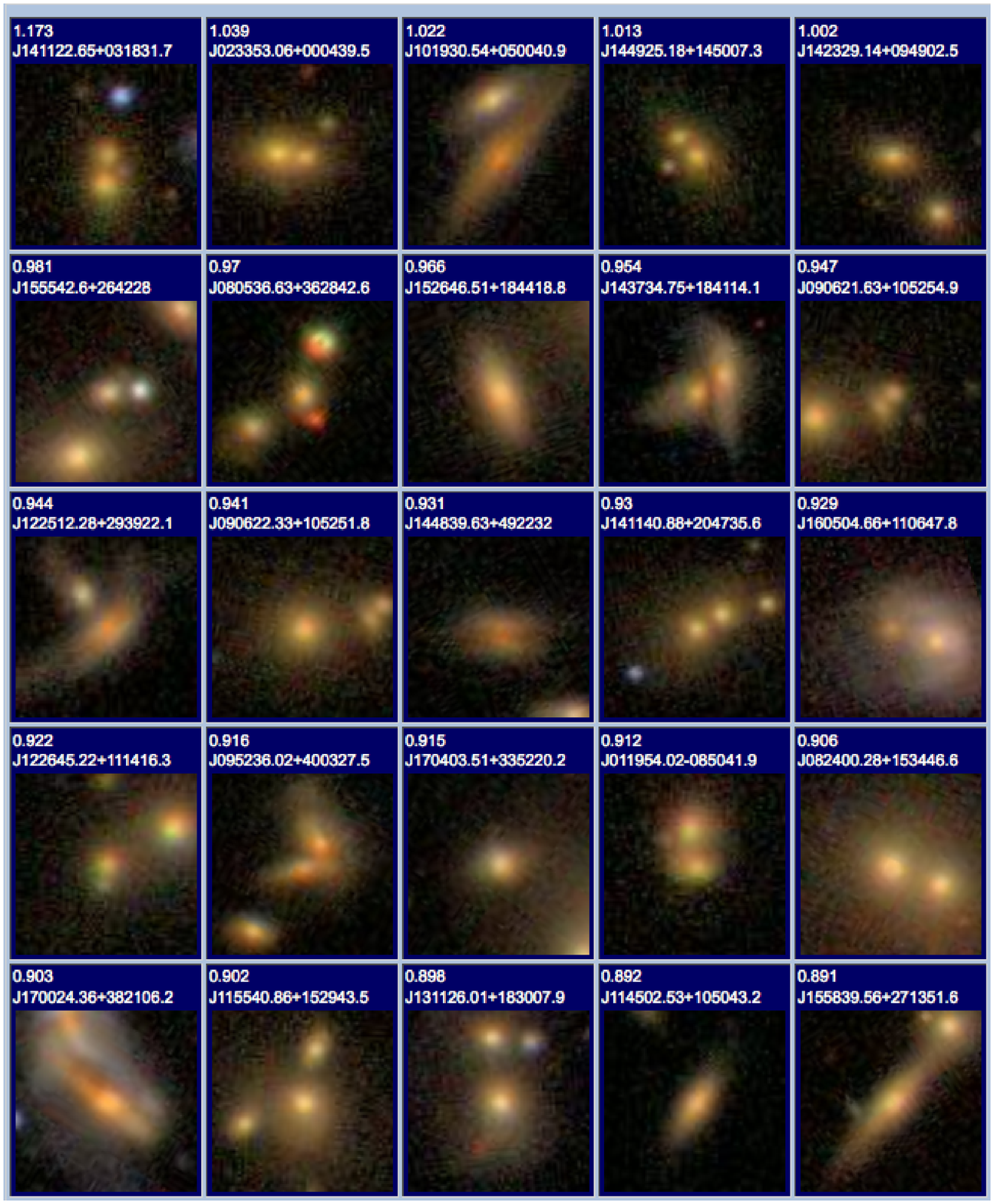}}}
\caption{Images of galaxies in close pairs ($r_p < 30$ \hkpc~ and 
$\Delta v < 200$ \kms) which have the bluest (left panel) and 
reddest (right panel) global colours. 
The colours are labelled in the top left of each image, 
and galaxies are sorted according to colour within each panel, with the 
most extreme at the upper left.
\label{figbluestreddest}}
\end{figure*}

Further insight into the nature of these 
extremely blue and extremely 
red galaxies can be gleaned by visual inspection of 
their images, as shown in Figure~\ref{figbluestreddest}.  
Clear morphological signs of interactions are seen 
within both sets of galaxies.  
There are obvious indications of dust
in some of the extremely red galaxies.  Some of this dust could have
been stirred up as a result of 
galaxy interactions \citep[e.g.,][]{geller06}.  However, 
in some cases, the dust is associated with edge-on discs; 
the ensuing reddening 
might be expected to be on the order of 0.1 mag in $g-r$ \citep{masters10}.
Nevertheless, edge-on discs would be expected to be present in equal 
measure in the control sample, so they are unlikely to be responsible for
differences between paired and control galaxies. 

Finally, we note that these images provide vivid evidence of the well-known 
Holmberg effect \citep{holmberg58}, in that galaxies within individual 
pairs tend to have colours which are similar to one another  
(i.e., there are few red-blue pairs).  

\section{Fibre Colours}\label{secfibre}

In Section~\ref{secfibredata}, we described our computation of rest-frame 
fibre colours, and the suitability of these colours as a probe of central
(rather than global) star formation.  In this section, we begin by 
analyzing the distribution of fibre colours in close, wide and projected pairs.
We then investigate the differences between fibre and global colours.

\subsection{The Distribution of Fibre Colours}

\begin{figure}
\rotatebox{0}{\resizebox{8.5cm}{!}
{\includegraphics{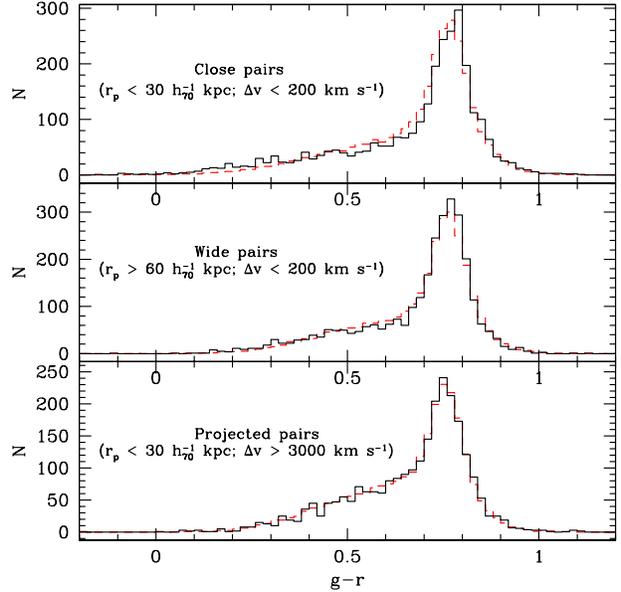}}}
\caption{Histograms of fibre $g-r$ are shown for pairs 
(solid black lines) 
and their associated control galaxies (dashed red lines). 
The lower plot is for projected pairs  
($r_p < 30$ \hkpc~ and $3000 < \Delta v < 10,000$ \kms), the middle plot 
is for wide pairs ($r_p > 60$ \hkpc~ and $\Delta v < 200$ \kms), and 
the upper plot is for close pairs ($r_p < 30$ \hkpc~ and $\Delta v < 200$ \kms).
\label{figgrfhist}}
\end{figure}

In Figure~\ref{figgrfhist}, 
we compare the fibre colours of paired and control galaxies, for 
close, wide and projected pairs.  We find a small but significant 
excess of extremely blue galaxies in close pairs, which is barely 
detectable in wide pairs and absent in projected pairs.  
On the other hand, we find no
excess of paired galaxies with extremely red fibre colours.  This 
is consistent with our hypothesis in \S~\ref{seceberphot} 
that residual problems with the 
photometry of crowded systems is responsible for the small population of
galaxies with extremely red global colours, since this effect should be much 
smaller when using fibre colours.

We also find a significant deficit of galaxies in close pairs 
with intermediate fibre colours (0.5 $< g-r <$ 0.75), 
and a corresponding excess of galaxies on the red half of the red sequence.  
This effect is smaller than was seen using global colours, particularly 
for wide pairs.  We again attribute this difference between paired and control 
galaxies to the higher density environments of paired galaxies.

\subsection{The Difference Between Fibre and Global Colours}

In the preceding subsection, we compared the distributions of fibre colours 
in close, wide and projected pairs to those found using global colours.  
While useful, this approach does not tell us how fibre and global colours
compare on a galaxy-by-galaxy basis.  If a subset of interacting galaxies 
experience centrally-concentrated bursts of star formation, we might 
expect to find evidence of this effect in the colour gradients of 
these galaxies.  In this section, we will use the difference between
fibre and global colour as a probe of this effect.  

\begin{figure}
\rotatebox{0}{\resizebox{8.5cm}{!}
{\includegraphics{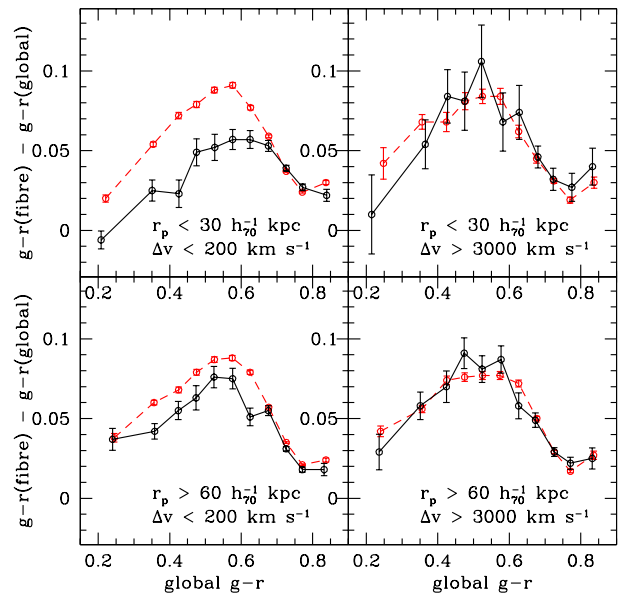}}}
\caption{The difference between fibre and global colours is plotted
versus global colour for four subsets of the pairs sample: 
close low velocity pairs (upper left panel), 
wide low velocity pairs (lower left panel), 
close projected pairs (upper right panel), and
wide projected pairs (lower right panel).
Paired galaxies are shown with black symbols and solid lines, 
while the associated 
control galaxies are shown with red symbols and dashed lines.  
Larger values of this colour difference correspond to redder centres.
\label{figfiboffall}}
\end{figure}

In Figure~\ref{figfiboffall}, we plot 
the difference between fibre and global colour 
versus global colour for four subsets of the pairs sample: 
close low velocity pairs (upper left panel), 
wide low velocity pairs (lower left panel), 
close projected pairs (upper right panel), and
wide projected pairs (lower right panel).
Overall, fibre colours are redder than global colours; 
i.e., $g-r$~(fibre) - $g-r$~(global) is positive.  This is as expected, 
given that bulges are generally redder than discs. 
The difference between fibre and global colours is 
largest for galaxies of intermediate colour.  The most obvious 
explanation is that galaxies of intermediate colour are the least likely 
to be either bulge-dominated or disc-dominated galaxies.  
In any case, we are primarily interested in how this index differs
between paired and control galaxies.

\begin{figure}
\rotatebox{0}{\resizebox{8.5cm}{!}
{\includegraphics{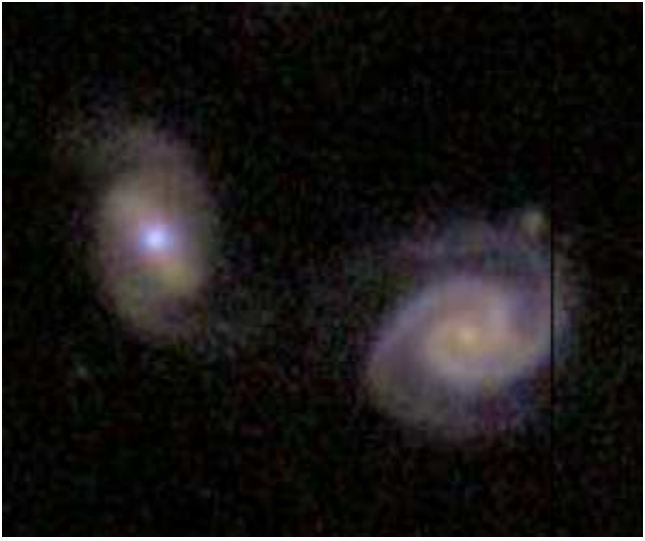}}}
\caption{An SDSS $gri$ image of a close galaxy pair which exhibits 
clear signs of a blue nucleus that may have been triggered by an interaction.  
The galaxy on the left (objID=587731513153159199)
has a fibre colour which is 0.22 mag bluer than its global colour.
This pair has a projected separation of 26 \hkpc~ and $\Delta v = 109$ \kms.
\label{figcentralsb}}
\end{figure}

Figure~\ref{figfiboffall} shows that, for red galaxies,
there is good agreement between the fibre-global colours 
of paired and control galaxies, for both wide and close pairs.  
Conversely, for blue galaxies, 
close low velocity pairs have fibre colours which are
typically 0.03 mag bluer than those of their associated 
control galaxies (upper left panel), 
with this difference dropping to 0.015 mag for wide low velocity 
pairs (lower left panel). 
This difference is absent in the projected 
pairs sample (right hand panels), 
confirming that crowding cannot be responsible for the offset.
This is a highly significant effect, and is our strongest indication yet 
that centrally triggered star formation is taking place in some of these
systems.  Figure~\ref{figcentralsb} gives a particularly striking example 
of a system in which morphological signs of an interaction accompany a 
relatively blue central colour.
The fact that the mean offset is largest in close pairs, but still
present in wide pairs, implies that the effects of this star formation 
on galaxy colours diminishes as galaxies move apart after close encounters, 
but that the effects persist long enough that they are still present 
in some widely separated pairs.  We explore this scenario further in
Section~\ref{secsb99}.

\section{Colour Offsets}\label{secoffset}

The increase in extremely blue galaxies in close pairs, combined 
with the relatively high {\it red} fraction of galaxies in 
close and wide pairs, 
demonstrates that galaxies in pairs can be either bluer or redder 
than their control counterparts, for different reasons.  
The former appears to be caused by 
ongoing galaxy-galaxy interactions, whereas the latter can be attributed
to the higher density environment occupied by paired galaxies, and 
may therefore be true for pre-interaction galaxy pairs too.  

In principle, these competing effects mean that some of the colour differences
between paired and control galaxies will cancel out when comparing their
colour distributions.  That is, only the {\it net changes} in the 
colour distributions will be detected.  
However, it may be possible to uncover more of the underlying colour 
differences by comparing, {\it on an individual 
basis}, the colour of every paired galaxy with the $\sim$ 12 galaxies in 
its associated control sample.  This method has the potential to detect 
much more of the underlying differences in galaxy colours, and may even allow 
us to identify which galaxies have had their colours changed the most 
as a result of ongoing/recent interactions.

To this end, 
we compute the {\it colour offset} for every paired galaxy in the sample.
We define colour offset as the colour of the paired galaxy minus
the mean colour of its associated ($\sim$ 12) control galaxies.
We also compute the colour offset of every control galaxy, by computing 
the difference between its colour and the mean colour of the remaining 
$\sim 11$ associated control galaxies.  We then compute the difference
between the colour offsets of paired and control galaxies (hereafter, 
we will refer to this quantity as $\Delta (g-r)$).  
If paired galaxy colours are drawn at random from the same parent population 
as control galaxies, we would expect to find $\Delta (g-r)$ to be zero, 
on average.

\begin{figure}
\rotatebox{0}{\resizebox{8.5cm}{!}
{\includegraphics{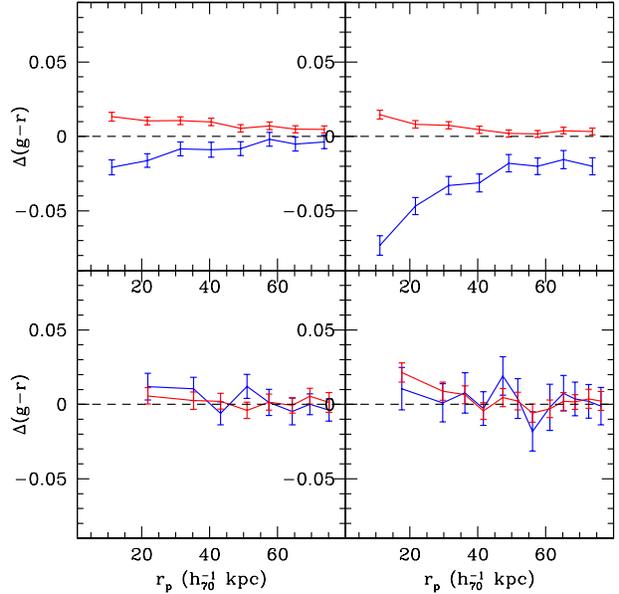}}}
\caption{The difference between the offset of paired and control galaxies
($\Delta g-r$) 
is plotted versus $r_p$ for low velocity pairs ($\Delta v < 200$ \kms; upper 
panels) and high velocity pairs ($\Delta v > 3000$ \kms; lower panels).  
The left hand panels refer to global colours, while the right hand
panels refer to fibre colours. 
Blue symbols refer to blue galaxies (those with global $g-r \leq 0.65$ for 
$M_r = -21$), and red symbols refer to red galaxies ($g-r > 0.65$ for 
$M_r = -21$).  In all plots, the dashed horizontal line
at $\Delta (g-r) = 0$ denotes the null hypothesis of no colour changes 
in paired galaxies.  
\label{figgroffsettrend}}
\end{figure}

Figure~\ref{figgroffsettrend} plots $\Delta (g-r)$ as a function of 
projected separation for low and high velocity pairs, and treats 
blue and red galaxies separately.  The left hand column of this figure
refers to global colours, and the right hand column refers to fibre colours.
For blue galaxies, a highly significant negative 
(blueward) fibre colour offset is found in low velocity pairs, changing
smoothly from 
about 0.02 mag for wide pairs to about 0.075 mag for the closest pairs.
At the smallest separations, this difference is significant 
at the 11$\sigma$ level.
This trend is much more modest in global colours, reaching $\sim$ 0.02 mag 
for the closest pairs.  This trend is absent in the high velocity (projected) 
pair sample, demonstrating that crowding errors cannot be responsible
for this effect.  

\begin{figure}
\rotatebox{0}{\resizebox{8.5cm}{!}
{\includegraphics{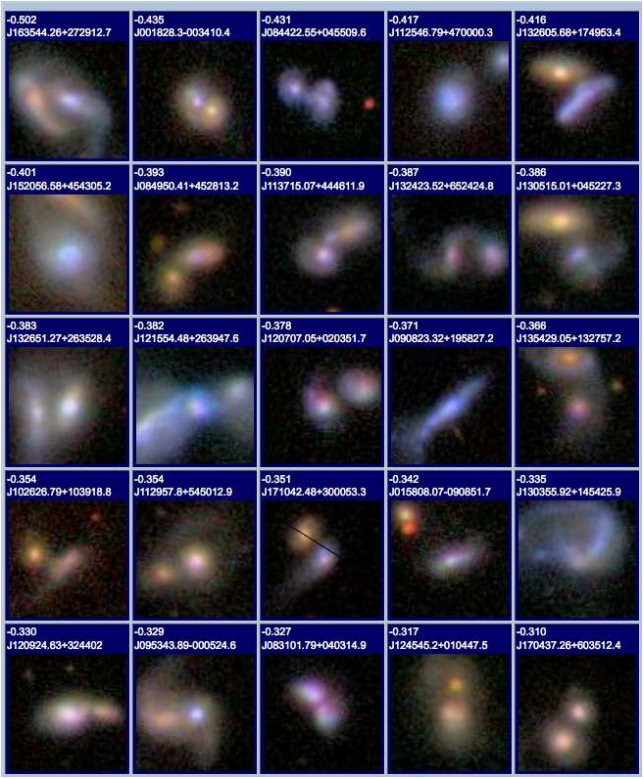}}}
\caption{Images of blue galaxies ($g-r < 0.65$ for $M_r = -21$) 
in very close pairs ($r_p < 15$ \hkpc~ and
$\Delta v < 200$ \kms) are shown for 
galaxies which have the greatest blueward offsets 
in fibre colours. 
The offsets are labelled in the top left of each image, 
and galaxies are sorted according to offset 
(greatest offset at the upper left).
\label{figoffblue}}
\end{figure}

Images of blue galaxies in very close pairs ($r_p < 15$ \hkpc) 
which have the greatest blueward offsets are presented in 
Figure~\ref{figoffblue}.  Many of these systems exhibit clear morphological
signs of interactions, along with indications of relatively blue central 
colours.  This demonstrates that our colour offset parameter does in fact
appear to be effective at identifying systems with atypical star forming 
properties (note that more than half of the galaxies in Figure~\ref{figoffblue} 
do not qualify as extremely blue, and therefore do not stand out when using
colour rather than colour offset).
Together, these results are consistent with the presence 
of central induced star formation which is strongest in the closest pairs 
and weaker (but still present) in wide pairs.

A much weaker trend is found for red galaxies in low velocity pairs, 
with an increase (reddening) in $\Delta (g-r)$ of up to 0.015 mag for the 
closest pair separations.  The size of this effect is 
the same in fibre and global colours, implying that the effect is global. 
However, it appears from Figure~\ref{figgroffsettrend} that a comparable
trend may also be present in the projected pairs sample; therefore, 
we are unable to 
rule out the possibility that crowding errors are responsible for this trend.

\subsection{Dependence on Projected Local Density}

\begin{figure}
\rotatebox{0}{\resizebox{8.5cm}{!}
{\includegraphics{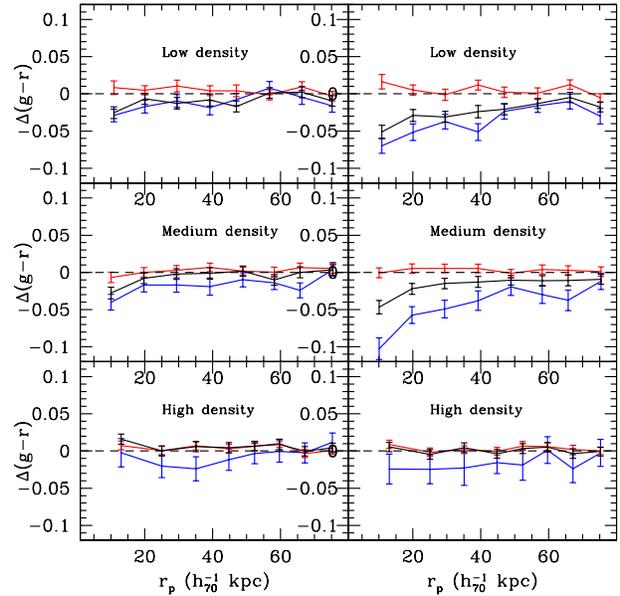}}}
\caption{The difference between the offset of paired and control galaxies
($\Delta g-r$) 
is plotted versus $r_p$ for low velocity pairs ($\Delta v < 200$ \kms)
for three tertiles in projected local density: low density (upper 
panels), medium density (middle panels), and high density (lower panels).
The left hand panels refer to global colours, while the right hand
panels refer to fibre colours. 
Blue symbols refer to blue galaxies (those with global $g-r \leq 0.65$ for 
$M_r = -21$), red symbols refer to red galaxies ($g-r > 0.65$ for 
$M_r = -21$), and black symbols refer to all galaxies (i.e., blue and red).  
In all plots, the dashed horizontal line
at $\Delta (g-r) = 0$ denotes the null hypothesis of no colour changes 
in paired galaxies.  
\label{figgrdensity}}
\end{figure}

In a related study, \citet{ellison10} found a small bluing of the bulges, 
and not the discs, of galaxies in close low velocity pairs.  
This effect was seen only at low densities, and was interpreted as
evidence of central triggered star formation.  Given the striking 
blueward fibre offsets seen in the upper right panel of 
Figure~\ref{figgroffsettrend}, we now investigate the dependence of
this effect on projected local density ($\Sigma$).  Measurements of 
$\Sigma$ are available for 57\% of our paired galaxies, as described
in Section~\ref{secrfrac}.
We subdivide our sample into three equal bins (tertiles) of $\Sigma$, 
and present the global and fibre colour offsets of each in 
Figure~\ref{figgrdensity}.  

This figure indicates clearly that the
blueward fibre offsets at small separations are driven by blue galaxies residing
in low and medium density environments, though a small blueward fibre offset 
is also detected in the highest density tertile.  Small blueward offsets in 
{\it global} colours (left hand column of Figure~\ref{figgrdensity}) are 
seen at small separations at low and medium densities, but not at high density.
The fact that the total (blue+red) population closely traces the red 
population in the high density regime and traces the blue population in 
the low density regime is consistent with the well known colour-density
relation, whereby the red fraction increases 
with density (see \S~\ref{secrfrac}).

Compared with figure 9 of \citet{ellison10}, we find a larger and more
significant difference between the colours of paired and control galaxies
at small separations, and we find that this difference extends further 
into the medium- to high-density regime.  We attribute this added sensitivity 
to triggered star formation to several factors: 
(1) our DR7 pairs sample is larger than the DR4 sample of \citet{ellison10}, 
(2) we have approximately 3 times as many control galaxies per paired
galaxy as \citet{ellison10}, 
(3) colour offsets are 
sensitive to colour differences on a galaxy-by-galaxy basis, and
(4) treating blue and red galaxies separately is effective in isolating
the effects of triggered star formation to the (blue) population of 
galaxies which is most susceptible to this process.  

\subsection{The Effects of Matching Control Galaxies on Density}

One significant difference between our study and several other 
close pair studies 
\citep[e.g.,][]{alonso06,perez09b,ellison10} is that we do not 
attempt to match our control sample to the projected local densities 
of paired galaxies.  As demonstrated in Section~\ref{secrfrac}, 
our resulting control sample is skewed to lower densities than the
pairs sample.  This has some implications for the interpretation of
differences between paired and control galaxies in our study.

We therefore investigate this issue by regenerating 
Figure~\ref{figgroffsettrend} 
using a control sample which {\it is} matched on local density.
This revised control sample is generated using the same methodology as
outlined in Section~\ref{seccontrol}, but now matching simultaneously 
on redshift, stellar mass and $\Sigma$.  However, since it is more difficult 
to find matches for paired galaxies in the highest density environments, 
we restrict our analysis to galaxies with log$(\Sigma) < 1.25$ (this excludes
12\% of galaxies in our paired sample).  We are able to find 3 control 
galaxies per paired galaxy.  

\begin{figure}
\rotatebox{0}{\resizebox{8.5cm}{!}
{\includegraphics{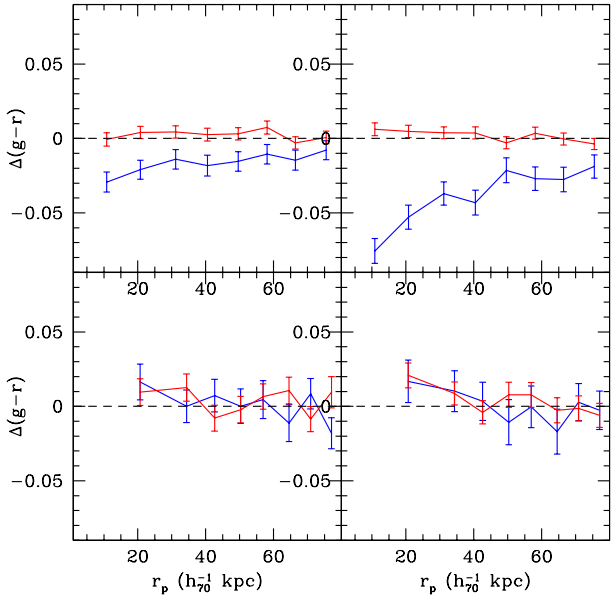}}}
\caption{This figure presents the same information as 
Figure~\ref{figgroffsettrend}, but restricts the analysis to galaxies with 
log$(\Sigma) < 1.25$ and uses a control sample which is matched on redshift, 
stellar mass {\it and} projected local density.
\label{figgrdensityc}}
\end{figure}

The results are given in Figure~\ref{figgrdensityc}.  Comparison with 
Figure~\ref{figgroffsettrend} reveals very similar trends in  
colours offsets.
With a density-matched control sample, there is a slight blueward shift 
in the global and fibre colour offsets of blue and red galaxies.
This small shift may result from the removal of galaxies in the highest density 
regime, since we know from Figure~\ref{figgrdensity} 
that galaxies in the highest density environments exhibit the smallest
blueward colour offsets.  Another factor which may contribute to this
shift is the fact that our revised control sample is no longer biased 
towards lower densities than paired galaxies; this would be expected to 
further accentuate the blueward colour offsets in pairs due to 
induced star formation.  We conclude that the main results 
of our study are unchanged if a density-matched control sample 
is used.

\section{A Simple Starburst Model}\label{secsb99}

The interpretation of the colours of interacting galaxies can be aided by 
employing model starbursts which predict how colours evolve during 
and after a starburst.  
We create a simplistic 
galaxy+starburst model by starting with two pre-existing galaxies 
(a red sequence galaxy and a blue cloud galaxy) 
and superimposing a model starburst upon each.  The goal here 
is to see how the colour of each galaxy changes with time as a 
result of the superimposed starburst.
For the colours of the starburst itself, we use an instantaneous 
Starburst99 model starburst 
\citep{leitherer99}, with $Z=0.020$, $\alpha = 3.30$, 
and $M_{\rm up} = 100 \msun$, 
converting from model $V-R$ colours to 
SDSS $g-r$.\footnote{We use the Lupton (2005) colour transformations 
found at http://www.sdss.org/dr5/algorithms/sdssUBVRITransform.html.}
We monitor these colours for 1 Gyr following the starburst.
For the pre-existing blue cloud galaxy, we use a Starburst99 model galaxy with 
a continuous star formation rate of $5 \msun$/year, and 
add the instantaneous starburst after
10 Gyr.  This galaxy has a $g-r$ colour of 0.43 immediately preceding
the starburst, and is therefore representative of the blue cloud galaxies
seen in Figure~\ref{figcmdcat}.  
For the pre-existing red sequence galaxy, we begin with an instantaneous 
starburst, and let it age for 10 Gyr, yielding a galaxy colour of 0.83, 
which lies within the red tail of the 
red sequence seen in Figure~\ref{figcmdcat}.  We then add the model
starburst and monitor how the galaxy's colour changes.  

\begin{figure}
\rotatebox{0}{\resizebox{8.5cm}{!}
{\includegraphics{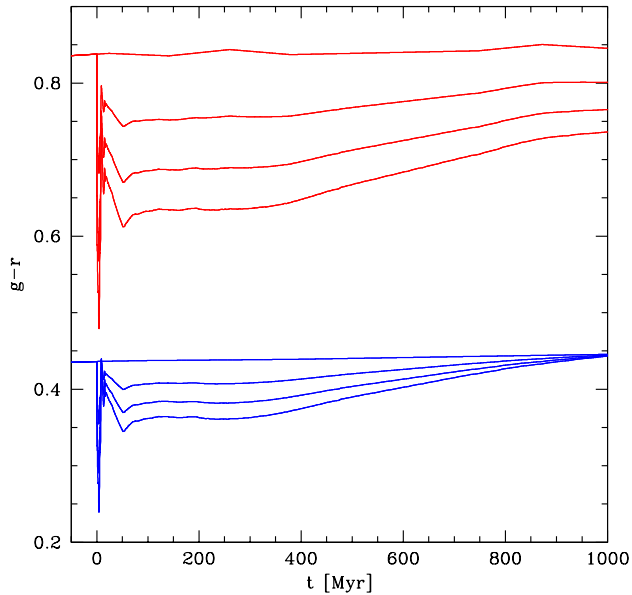}}}
\caption{The colour evolution due to starbursts is modelled by 
adding Starburst99 model starbursts to a pre-existing red sequence 
and blue cloud galaxy.  
The uppermost red (blue) line shows the colour evolution of
the red sequence (blue cloud) galaxy without a starburst.  After adding  
starbursts at time $t$=0 with burst strengths (by stellar mass) 
of 10\%, 20\%, and 30\% 
(top to bottom), the colour evolution of the resulting galaxies is followed 
for 1 Gyr.  Both galaxies become bluer initially, followed by a 
gradual reddening.  The colour change is about three times larger for the 
red sequence galaxy, due to the fact that the starburst is much bluer than
the pre-starburst galaxy.  After 1 Gyr, the late-type galaxy has returned to 
its pre-burst colour, whereas the early-type galaxy is still noticeably 
bluer.
\label{figgrt}}
\end{figure}

The evolution in colours of these galaxy+starburst models
is shown in Figure~\ref{figgrt}, for burst strengths (by stellar mass)
of 10\%, 20\%, and 30\%.
This figure shows that a 20\% starburst 
within a pre-existing red sequence 
galaxy will cause the galaxy to become bluer
by $\sim 0.15$ mag, 
with this offset persisting for $\sim 400$ Myr before gradual 
reddening begins.  Even at
1 Gyr after the starburst, this galaxy is considerably bluer than 
it was initially.  Conversely, for a starburst that occurs in a blue cloud
galaxy, the galaxy will become $\sim 0.05$ mag bluer
for about 400 Myr, and will return to its pre-starburst colour after 1 Gyr.  

To compare with our results from the previous section, we note that 
the difference in colour offset between paired and control galaxies 
(see Figure~\ref{figgroffsettrend}) should be analogous to the offsets between 
our model starburst galaxies and their pre-existing counterparts 
(Figure~\ref{figgrt}), if all galaxies in our close pair sample were 
undergoing interactions with induced star formation.  In reality, of course, 
not all of the galaxies in close pairs can be undergoing interactions, as 
some will not be close in three dimensions (i.e., interlopers), some will 
be approaching one another and therefore will not yet have had a close 
encounter, and others may be undergoing interactions without triggered
star formation.  Therefore, we would expect the mean colour offsets in our 
close pairs sample to be smaller than our model predictions, perhaps by 
a factor of a few.

In Figure~\ref{figgroffsettrend}, we found that blue galaxies in close pairs 
have mean global colour 
offsets which are $\sim$ 0.02 mag bluer than their control galaxies.  This 
is smaller than the $\sim$ 0.05 mag offset predicted for a 20\% starburst
in a blue galaxy (Figure~\ref{figgrt}), but consistent within the hypothesis
that 40\% of galaxies in close pairs are experiencing induced starbursts
of this strength.  We find a much higher offset of 0.075 mag in fibre 
(rather than global) colours.   
This makes sense if the fractional starbursts in the central 
regions of these galaxies are substantially higher than 20\%, as would be 
the case if most of the induced star formation is centrally concentrated, 
as implied by Figure~\ref{figfiboffall}.  
Finally, the blueward offsets we detect decrease markedly going from close 
to wide pairs.  This is consistent with the aging and subsequent reddening
of our models in Figure~\ref{figgrt}, and implies that we are seeing 
starbursts age as galaxies in close pairs separate following 
close peri-centre encounters.  The fact that this offset is still visible 
at the largest separations probed (80 \hkpc) is consistent with a simple 
calculation showing that a pair of galaxies separating at 200 \kms~ in
the plane of the sky will take $\sim 400$ Myr to reach $r_p \sim 80$ \hkpc.

For red galaxies, we find a small {\it redward} offset in close pairs 
(Figure~\ref{figgroffsettrend}), which may be due to residual errors
with the photometry.  We see no indications of the potentially 
large blue colour offsets
that are predicted by our model galaxy+starburst for red galaxies.  
This appears to indicate that starbursts of order 10-30\% are not 
commonly found in red galaxies.  While it is obvious that sufficiently 
large offsets would move red sequence galaxies into the blue cloud, 
the absence of even a small blueward offset for typical red galaxies 
appears to rule out this scenario.  
This finding is not unexpected, however, 
given that red sequence galaxies are generally depleted in gas, and 
therefore less capable of triggered star formation.  
While deep imaging reveals that gas-poor (early type) galaxies 
show frequent signs of tidal disturbances, these signs of interactions 
are not generally accompanied by star formation 
\citep{tal09,ellison10,kaviraj10}.  \citet{woods07} find a correlation 
between the specific SFR and pair separation for galaxies in blue pairs, 
but not in red pairs, confirming this interpretation.
Moreover, 
while few red mergers contain significant amounts of dust \citep{whitaker08}, 
\citet{gallazzi09} find that, in intermediate to high density environments
(within which many of our close pairs lie), galaxies which have had most
of their star formation suppressed tend to have their remaining star 
formation obscured.  
In summary, it appears that red sequence galaxies in close pairs 
exhibit little in terms of induced star formation in the optical, 
with few (if any) unobscured starbursts with large burst strengths.

\section{Conclusions}\label{secconclusions}

We have compiled a large, well-defined sample of 21,347 SDSS DR7 galaxies 
in pairs with $r_p < 80$ \hkpc, $\Delta v < 10,000$ \kms, and stellar mass ratios 
between 0.1 and 10.  We have measured high quality $g-r$ global colours 
for each galaxy, and have acquired their central (fibre) colours.
We have also created a very large control sample which is matched to the
pairs sample in stellar mass and redshift, with $\sim 12$ control galaxies
associated with every paired galaxy.
By comparing galaxies in close pairs with their control samples, with 
wider separation pairs, and with close projected pairs (i.e., interlopers), 
we have been able to distinguish
trends which are associated with interactions from those which are due to 
differences in environment or those which result from photometric errors
due to crowding.  Our findings can be summarized as follows:
\begin{enumerate}
\item 60\% of galaxies in close and wide pairs are classified as red, 
compared with 56\% of control galaxies.  These paired galaxies are 
found in higher density environments than their controls.  
We interpret these results as an indication that galaxies which are
involved in interactions are preferentially red before the interactions
start, due to the older stellar populations which are present in higher 
density environments.
\item Galaxy-galaxy interactions make blue galaxies in close pairs 
($r_p < 30$ \hkpc~ and $\Delta v <$ 200 \kms) 
bluer by an average of 0.075 mag in fibre colours, and
0.02 mag in global colours.  These colour offsets are diminished but 
still detectable out to pair separations of at least 80 \hkpc, and
are strongest at low and medium projected local densities.
\item The fraction of extremely blue galaxies rises from about 3\% for 
wide pairs to 8\% for close pairs.  The use of projected pairs (interlopers) 
and alternate colour measurements confirm that this effect is not due
to photometric errors.
\item Galaxy-galaxy interactions appear to 
have little (and perhaps no) effect on
the optical colours of red galaxies in pairs.  The slight reddening we detect 
at small separations (up to 0.015 mag in global and fibre colours) 
could be due to dust obscuration, but the rapid decline with pair
separation and the absence of any difference between global and
fibre colours instead suggest that 
residual problems with crowded-field photometry may be responsible.
\item Unlike previous studies of close pairs, we find no significant excess
($< 1\%$) of extremely red galaxies in close pairs.  
We demonstrate that this is due
to our improved photometry in crowded systems, 
given that we do find a strong excess ($\sim 6\%$) if we replace our 
{\sc GIM2D} colours with Petrosian or modelMag colours from the SDSS database.
\item At a fixed global colour, blue cloud galaxies in close pairs have bluer 
fibre colours than control galaxies, with this difference decreasing as
pair separation increases.  No such difference is found for red galaxies
at any separations.
\item Our simple starburst+galaxy model predicts that a 20\% induced starburst
should make a blue cloud (red sequence) galaxy bluer by about 
0.05 mag (0.15 mag), and should persist for $\sim$ 400 Myr before starting
to diminish.  Our observed colour offsets indicate that starbursts such as
this are in fact found in blue cloud galaxies in pairs, 
but are absent in red sequence galaxies in pairs.
\end{enumerate}

Together, these results provide further evidence that gas-rich galaxies 
in close pairs undergo induced star formation during close peri-centre 
passages, with the starburst then aging as the galaxies move apart from
one another.  Fibre colours confirm that this star formation is
centrally concentrated, and measurements of projected local density show that 
this process occurs primarily in low- to medium-density 
environments.  We find no evidence from optical colours 
for such induced star formation in red sequence galaxies, thereby confirming
that any such star formation is likely to be obscured.
We refute earlier claims of a substantial excess of 
extremely red galaxies in close pairs.

\section*{Acknowledgments} 

We thank the anonymous referee for insightful comments which significantly 
improved this paper.
We are grateful to Anja von der Linden and the MPA/JHU group for
access to their data products and catalogues (maintained by Jarle
Brinchmann at http://www.mpa-garching.mpg.de/SDSS/).
We thank Ivan Baldry for making his DR7 projected local density measurements 
publicly available.
SLE and DRP gratefully acknowledge the receipt of NSERC Discovery Grants which
funded this research.

The SDSS is managed by the Astrophysical Research Consortium for the 
Participating Institutions. The Participating Institutions are the 
American Museum of Natural History, Astrophysical Institute Potsdam, 
University of Basel, University of Cambridge, Case Western Reserve University, 
University of Chicago, Drexel University, Fermilab, the Institute for 
Advanced Study, the Japan Participation Group, Johns Hopkins University, 
the Joint Institute for Nuclear Astrophysics, the Kavli Institute for 
Particle Astrophysics and Cosmology, the Korean Scientist Group, 
the Chinese Academy of Sciences (LAMOST), Los Alamos National Laboratory, 
the Max-Planck-Institute for Astronomy (MPIA), the Max-Planck-Institute for 
Astrophysics (MPA), New Mexico State University, Ohio State University, 
University of Pittsburgh, University of Portsmouth, Princeton University, 
the United States Naval Observatory, and the University of Washington.

\end{document}